\RequirePackage{fix-cm}
\documentclass{svjour3}                     
\smartqed  
\usepackage{graphicx}
\usepackage{natbib}
\usepackage{color}
\usepackage{times}
\usepackage{latexsym}
\usepackage{enumerate}
\usepackage{epsfig}
\usepackage{amsmath}
\usepackage{amssymb}
\usepackage{amsfonts}
\usepackage[english]{babel}
\usepackage{graphicx}
\usepackage{graphics}
\usepackage{psfrag}
\usepackage{amsbsy}
\usepackage{amsmath}
\usepackage{textcomp}

\usepackage{mathptmx}       
\usepackage{helvet}         
\usepackage{courier}        
\usepackage{type1cm}        
\usepackage{makeidx}         
\usepackage{graphicx}        
\usepackage{multicol}        
\usepackage[bottom]{footmisc}

\usepackage{amsmath}
\usepackage{amsfonts}

\usepackage{amsmath}
\usepackage{amsfonts}

\def\ba{\mathbf{a}}

\def\be{\mathbf{e}}
\def\bx{\mathbf{x}}

\def\bz{\mathbf{z}}

\def\bu{\mathbf{u}}

\def\bA{\mathbf{A}}
\def\bH{\mathbf{H}}

\def\bX{\mathbf{X}}

\def\bB{\mathbf{B}}
\def\bD{\mathbf{D}}
\def\bS{\mathbf{S}}
\def\bI{\mathbf{I}}

\def\bU{\mathbf{U}}
\def\bV{\mathbf{V}}
\def\bZ{\mathbf{Z}}
\def\b0{\mathbf{0}}
\def\b1{\mathbf{1}}

\def\cL{\mathcal{L}}
\def\cN{\mathcal{N}}

\def\mR{\mathbb{R}}

\def\EE{\mathbb{E}}

\def\bmu{\mbox{\boldmath $\mu$}}

\def\balpha{\mbox{\boldmath $\alpha$}}

\def\bgamma{\mbox{\boldmath $\gamma$}}

\def\btheta{\mbox{\boldmath $\theta$}}

\def\bSigma{\mbox{\boldmath $\Sigma$}}
\def\bDelta{\mbox{\boldmath $\Delta$}}
\def\bGamma{\mbox{\boldmath $\Gamma$}}
\def\bLambda{\mbox{\boldmath $\Lambda$}}

\def\bPsi{\mbox{\boldmath $\Psi$}}
\def\bTheta{\mbox{\boldmath $\Theta$}}

\def\sbZ{\underset{\widetilde{}}{\bZ}}
\def\sbX{{\underset{\sim}{\bX}}}

\def\diag{\mathrm{diag}}
\def\tr{\mathrm{tr}}

\begin{document}

\title{Maximum likelihood estimation in constrained parameter spaces for mixtures of factor analyzers}
%
%
\author{Francesca Greselin \and Salvatore Ingrassia}
\institute{Francesca Greselin \at Department of Statistics and  Quantitative Methods\\  Milano-Bicocca University\\
Via Bicocca degliArcimboldi 8 - 20126 Milano (Italy). \email{francesca.greselin@unimib.it}
\and 
Salvatore Ingrassia\at Department of Economics and Business\\ University of Catania\\
Corso Italia 55, - Catania (Italy). \email{s.ingrassia@unict.it}
}

\date{Received: date / Accepted: date}
%
%
\maketitle
\begin{abstract}
Mixtures of factor analyzers are becoming more and more popular in the area of model based clustering of high-dimensional data. According to the likelihood approach in data modeling, it is well known that the  
unconstrained log-likelihood function may present spurious  maxima and  singularities and this is due to specific patterns of the estimated covariance structure, when their determinant approaches 0.  To reduce such drawbacks, in this paper we introduce  a procedure for the parameter estimation of mixtures of factor analyzers, which maximizes the likelihood function in a constrained parameter space.
We then analyze and measure its performance, compared to the usual non-constrained approach, via some simulations and applications to real data sets.
%
%

\end{abstract}
%


\section{Introduction and motivation}
Finite mixture distributions have been receiving a growing interest in statistical modeling. Their central role is mainly due to their double nature: they combine the flexibility of non-parametric models with the  strong and useful mathematical properties of parametric models.  According to this approach, when we know that a sample of observations has been drawn from different populations, we assume a specific distributional form in each of the underlying populations. The purpose is to decompose the sample into its mixture components, which, for quantitative data, are usually modeled as a multivariate Gaussian distribution, and to estimate parameters. The assumption of underlying normality, besides the elegant analytic properties, allows also to employ the EM algorithm for the ML estimation of the parameters.
On the other side, when considering a large number of observed variables, Gaussian mixture models can provide an over-parameterized solution as, besides the mixing weights, it is required to estimate the mean vector and the covariance matrix for each component \citep{Peel:McLa:2000}. As a consequence, we observe at the same time an undue load of computationally intensive procedures for the estimation. 

This is the reason why a number of strategies have been introduced in the literature to avoid over-parameterized solutions.
Among the various proposal, some authors developed methodologies for variable selection (see, f.i.,  \citet{Liu:2003} and \citet{Hoff:2005} in the Bayesian framework, \citet{Pan:Shen:2007} and  \citet{Raft:Dean:Vari:2006} in the frequentist one). They further motivate their approach from the observation that the presence of non-informative variables can be strongly misleading for some clustering methods.
With the same purpose of parsimony, but a completely different approach, \citet{Banf:Raft:mode:1993} devised a methodology to identify common patterns among the  component-covariance matrices; their proposal arose a great attention in the literature.  Along a slightly different line of thinking,  \citet{Ghah:Hilt:1997} and \citet{McLa:Peel:Bean:2003} proposed to employ latent variables to perform dimensional reduction in each component, starting from the consideration that in many phenomena some few unobserved features could be explained by the many observed ones.

In this paper we address mixtures of factor analyzers by assuming that the data have been generated by a linear factor model with latent variables modeled as Gaussian mixtures. Our purpose is to improve the performances of the EM algorithm, by facing with some of its issues and giving practical recipes to overcome them.  It is well known that the EM algorithm  generates a sequence of estimates, starting from an initial guess, so that the corresponding sequence of the log-likelihood values is not decreasing.  However, the convergence toward the MLE is not guaranteed, because the log-likelihood is unbounded and presents local maxima. Another critical point is that the parameter estimates as well as the convergence of the whole estimation process may be affected by the starting values (see, f.i., \citet{McKr:theE:2007}) so that the final estimate crucially depends on the initial guess.
This issue has been investigated by many authors, starting from the seminal paper of \cite{Redn:Walk:1984}. Along the lines of \citep{Ingr:2004}, in this paper we  introduce and implement a procedure for the parameters estimation of mixtures of factor analyzers, which maximizes the likelihood function in a constrained parameter space, having no singularities and a reduced number of spurious local maxima. We then analyze and measure its performance, compared to the usual non-constrained approach.

We have organized the rest of the paper as follows. In Section \ref{sec:GaussianFA} we summarize main ideas about Gaussian Mixtures of Factor Analyzer model; 
in Section \ref{sec:The AECM algorithm} we provide fairly extensive notes concerning  the likelihood function and the AECM algorithm.
Some well known considerations \citep{Hath:Acon:1985}  related to spurious maximizers and  singularities in the EM algorithm are recalled in Section \ref{ConstrainedML}, and motivate our proposal to  introduce constraints on factor analyzers. Further, we give a detailed methodology to implement such constraints into the EM algorithm. 
In Section \ref{sec:numerical results} we show and discuss  the improved performance of our procedure, on the ground of some numerical results based on both simulated and real data. Section \ref{sec:concluding} contains concluding notes and provides ideas for future research.

\section{The Gaussian Mixture of Factor analyzers}\label{sec:GaussianFA}
%
Within the Gaussian Mixture (GM) model-based approach to density estimation and clustering, the density of the $d$-dimensional random variable $\bX$ of interest is modelled as a mixture of a number, say $G$, of multivariate normal densities in some unknown proportions $\pi_1,\ldots \pi_G$. That is, each data point is taken to be a realization of the mixture probability density function,
\begin{equation}
f(\bx;\btheta)=\sum_{g=1}^G \pi_g \phi_d(\bx;\bmu_g,\bSigma_g)\label{mixt-gaussian}
\end{equation}
where $\phi_d(\bx;\mu,\bSigma)$ denotes the $d$-variate normal density function with mean $\bmu$ and covariance matrix $\bSigma$. Here the vector $\btheta_{GM}(d,G)$ of unknown parameters consists of the $(G-1)$ mixing proportions $\pi_g$, the $G \times d$ elements of the component means $\mu_g$, and the ${1 \over 2} G d (d+1)  $ distinct elements of the component-covariance matrices $\bSigma_g$.
Therefore, the $G$-component normal mixture model (\ref{mixt-gaussian}) with unrestricted component-covariance matrices is a highly parametrized  model. We crucially need some method for parsimonious parametrization of the matrices  $\bSigma_g$, because they requires $O(d^2)$ parameters.
Among the various proposals for dimensionality reduction, we are interested here in considering Mixtures of Gaussian Factor Analyzers (MGFA), which allows to explain data by explicitly modeling correlations between variables in multivariate observations. We postulate a finite mixture of linear sub-models for the distribution of the full observation vector $\bX$, given the (unobservable) factors $\bU$. That is we can provide a local dimensionality reduction method by assuming that the distribution of the observation  $\bX_i$  can be given as
\begin{equation}
\bX_i=\bmu_g+\bLambda_g\bU_{ig}+\be_{ig} \quad\textrm{with probability }\quad \pi_g \: (g=1,\ldots,G) \quad \textrm{for} \,\, i=1,\ldots,n, \label{factor_an}
\end{equation}
where $\bLambda_g$ is a $d \times q$ matrix of \textit{factor loadings}, the  \textit{factors} $\bU_{1g},\ldots, \bU_{ng}$ are $\cN(\mathbf{0},\bI_q)$ distributed independently of the  \textit{errors} $\be_{ig}$, which are  independently $\cN(\mathbf{0},\bPsi_g)$  distributed, and $\bPsi_g$ is a $d \times d$ diagonal matrix $(g=1,\ldots,G)$. We suppose that $q<d$, which means that  $q$ unobservable factors are jointly explaining the $d$ observable features of the statistical units.
Under these assumptions, the mixture of factor analyzers model is given by (\ref{mixt-gaussian}), where the $g$-th component-covariance matrix $\bSigma_g$ has the form
\begin{equation}
\bSigma_g=\bLambda_g \bLambda'_g+\bPsi_g \quad (g=1,\ldots,G). \label{Sigmag}
\end{equation}
The parameter vector $\btheta_{MGFA}(d,q,G)$ now consists of the elements of the component means $\bmu_g$, the $\bLambda_g$, and the $\bPsi_g$, along with the mixing proportions $\pi_g$ $(g=1,\ldots,G-1)$, on putting $\pi_G=1-\sum_{i=1}^{G-1}\pi_g$.
Note that in the case of $q>1$, there is an infinity of choices for $\bLambda_g$, since model (\ref{factor_an}) is still satisfied if we replace $\bLambda_g$ by  $\bLambda_g \bH'$, where $\bH$ is any orthogonal matrix of order $q$. As $q(q-1)/2$ constraints are needed for $\bLambda_g$ to be uniquely defined, the number of free parameters, for each component of the mixture, is 
\[ dq+d-{1 \over 2 } q (q-1) . \]


Comparing the two approaches and willing now to measure the gained parsimony when we use mixtures of factor analyzers, with respect to the more usual gaussian mixtures,  and denoting by $|\btheta_{CovGM}(d,G)|$  and  $|\btheta_{CovMGFA}(d,q,G)|$, the number of the estimated parameters for the covariance matrices in the GM and MGFA models, respectively, we have to choose  values of $q$ such that the following quantity $P$ is positive
\[P=|\btheta_{CovGM}(d,G)|-|\btheta_{CovMGFA}(d,q,G)|={ G \over 2 } d (d+1)-G[dq-d+{1 \over 2 } q (q-1)]\]
i.e.:
\[P={ G \over 2 } [(d-q)^2-(d+q) ] . \]
This is the only requirement for parsimony.  Now, we can express  the relative reduction $RR(d,q,G)=RR(d,q)$ given by 
 \begin{align*}
RR(d,q) &=\frac{|\btheta_{CovGM}(d,G)|-|\btheta_{CovGMFA}(d,q,G)|}{|\btheta_{CovGM}(d,G)|} = \frac{(d-q)^2-(d+q)}{d(d+1)}.
\end{align*}
 In Table \ref{tab:RR} we report the relative reduction, in term of lower number of  estimated parameters for the covariance matrices in the MGFA models, with respect to the GM models.

 \begin{scriptsize}
 
\begin{table}[h!]
\begin{center}
\caption{Relative reduction $RR(d,q)$}\label{tab:RR}
\begin{tabular}{ c | ccc ccc ccc ccc ccc}
  \hline    
  $q|d$	&1		&2		&3		&4		&5		&6	&7 &8	&9	&10	&11	&12	&13	&14	&15\\
   
 1	&-	&-	&-	&0.20	&0.33	&0.43	&0.50	&0.56	&0.60	&0.64	&0.67	&0.69	&0.71	&0.73	&0.75\\
2	&-	&-	&-	&-		&0.07	&0.19	&0.29	&0.36	&0.42	&0.47	&0.52	&0.55	&0.58	&0.61	&0.63\\
3	&-	&-	&-	&-		&-		&-		&0.11	&0.19	&0.27	&0.33	&0.38	&0.42	&0.46	&0.50	&0.53\\
4	&-	&-	&-	&-		&-		&-		&-		&0.06	&0.13	&0.20	&0.26	&0.31	&0.35	&0.39	&0.43\\
5	&-	&-	&-	&-		&-		&-		&-		&-		&0.02	&0.09	&0.15	&0.21	&0.25	&0.30	&0.33   \\ 
   \hline  
\end{tabular}
\end{center}
\end{table}
\end{scriptsize}
The relative reduction represents the extent to which the factor model offers a simpler interpretation for the behaviour of $\bx$ than the alternative assumption given by the gaussian mixture model.

\section{The likelihood function and the EM algorithm for MGFA}\label{sec:The AECM algorithm}
In this section we summarize the main steps of the EM algorithm for mixtures of Factor analyzers, see e.g. \cite{McLa:Peel:fini:2000} for details.

Let $\sbX=(\bx_1, \ldots, \bx_n)$ be a sample of size $n$ from density (\ref{mixt-gaussian}), and 
let  $\bx_i$ ($i=1, \ldots, n$) denotes the realization of $\bX_i$ in (\ref{factor_an}).
For given data $\sbX$, parameters in (\ref{mixt-gaussian}) can be estimated according to the likelihood approach via the EM
algorithm, where the likelihood function is given by:
\begin{align*} 
L(\btheta;  \sbX) & = \prod_{i=1}^n  \left\{ \sum_{g=1}^G  \phi_d(\bx_i; \bmu_g,\bSigma_g)  \, \pi_g \right\} = 
\prod_{i=1}^n  \left\{ \sum_{g=1}^G  \phi_d(\bx_i; \bmu_g,\bLambda_g,\bPsi_g)  \, \pi_g \right\} \, , 
\end{align*}
where we set $\bSigma_g = \bLambda_g \bLambda'_g + \bPsi_g$ ($g=1, \ldots, G$). Consider the augmented data $\{ (\bx_i, \bu_{ig}, \bz_i), \, i=1, \ldots, n \}$,
where $\bz_i = (z_{i1}, \ldots, z_{ig})'$, with  $z_{ig}=1$ if $\bx_i$ comes from the $g$-th population and $z_{ig}=0$ otherwise. 
Then, the complete-data likelihood function  can be written in the form:
\begin{equation}
L_c(\btheta; \sbX) = \prod_{i=1}^n\prod_{g=1}^G\left[ \phi_d\left(\bx_i|\bu_i;\bmu_g,\boldsymbol{\Lambda}_g,\boldsymbol{\Psi}_g\right)\phi_q(\bu_{ig}) \pi_g \right]^{z_{ig}}. \label{eq:complete-data log-likelihood gen} 
\end{equation}
In particular, due to the factor structure of the model, see \citet{Meng:VanD:TheE:1997}, we have to consider the  alternating expectation-conditional maximization (AECM) algorithm.
Such a procedure  is an extension of the EM algorithm that uses different specifications of missing data at each stage. The idea is to  partition $\btheta=(\btheta'_1, \btheta'_2)'$ in 
such a way that $L(\btheta; \sbX)$ is easy to maximize for $\btheta_1$ given $\btheta_2$ and vice versa. Then, we can iterate between these two conditional maximizations until convergence. 
In this case $\btheta_1=\{ \pi_g, \bmu_g, \,  g=1, \ldots, G \}$ where the missing data are the unobserved group labels $\sbZ=(\bz'_1, \ldots, \bz'_n)$, 
and the second part of the parameters vector is given by $\btheta_2=\{ (\bLambda_g, \bPsi_g),   \, g=1, \ldots, G \}$ where the missing data are the group labels $\bZ$ and the unobserved latent factors $\bU=(\bU_{11}, \ldots, \bU_{nG})$.
Hence, the application of the AECM algorithm consists of two cycles, and there is one E-step and one CM-step alternatively considering  $\btheta_1$ and $\btheta_2$ in each pair of cycles.
\paragraph{First Cycle.} Here it is $\btheta_1=\{ \pi_g, \bmu_g, \,  g=1, \ldots, G \}$ where the missing data are the unobserved group labels $\bZ=(\bz'_1, \ldots, \bz'_n)$. 
The complete data likelihood is
\begin{align}
L_{c1}(\btheta_1) & =\prod_{i=1}^n\prod_{g=1}^G\left[ \phi_d\left(\boldsymbol{\bx}_i;\bmu_g,\bSigma_g\right)\pi_g \right]^{z_{ig}} .
\label{eq:complete-data likelihood1}
\end{align}
%
The E-step on the first cycle on the $(k+1)$-th iteration requires the calculation of 
$Q_1(\btheta_1; \btheta^{(k)}) = \EE_{\btheta^{(k)}} \{\cL_{c} (\btheta_1) | \sbX \}$
which is the  expected complete-data log-likelihood given the data $\sbX$ and using the current estimate $\btheta^{(k)}$ for $\btheta$. In practice it requires calculating
$\EE_{\btheta^{(k)}} \{Z_{ig}| \sbX \}$ and usual computations show that this step is achieved by replacing each $z_{ig}$ by its current conditional expectation given the observed data $\bx_i$, that is we replace $z_{ig}$ by $z_{ig}^{(k+1/2)}$, where
\begin{equation}
z_{ig}^{(k+1)} =\frac{\phi_d\left(\bx_i|\bmu_g^{(k)},\bLambda_g^{(k)},\bPsi^{(k)}_g\right) \pi_g^{(k)} }{\sum_{j=1}^G \phi_d \left(\bx_i|\bmu_j^{(k)},\bLambda_j^{(k)},\bPsi_j^{(k)}\right) \pi_j^{(k)}} .
\end{equation}
On the M-step, the maximization of this complete-data log-likelihood yields
\begin{align*}
\pi_g^{(k+1)} & =\frac{\sum_{i=1}^n z_{ig}^{(k+1)}}{n}
\\
\bmu_g^{(k+1)} &=\frac{1}{n_g} \sum_{i=1}^n z_{ig}^{(k+1)} \bx_i 
\end{align*}
where $n_g^{(k+1)}=\sum_{i=1}^n z_{ig}^{(k+1)}$. According to notation in \citet{McLa:Peel:fini:2000}, we set $\btheta^{(k+1/2)}=(\btheta_1^{(k+1)'}, \btheta_2^{(k)'})'$.
%
\paragraph{Second Cycle.} Here it is $\btheta_2=\{ \bSigma_g, \,   g=1, \ldots, G \}= \{  (\bLambda_g$, $\bPsi_g), \,  g=1, \ldots, G \}$ where the missing data are the unobserved group labels $\bZ$ and the latent factors $\bU$.
Therefore, the complete data likelihood is
\begin{align}
L_{c2}(\btheta_2) & =\prod_{i=1}^n\prod_{g=1}^G\left[\phi_d\left(\bx_i|\bu_{ig};\bmu_g^{(k+1)},\bSigma_g\right)\phi_q\left(\bu_{ig}\right) \pi_g^{(k+1)} \right]^{z_{ig}} \nonumber \\
& = \prod_{i=1}^n\prod_{g=1}^G\left[\phi_d\left(\bx_i|\bu_{ig};\bmu_g^{(k+1)},\bLambda_g,\bPsi_g\right) \phi_q\left(\bu_{ig}\right) \pi_g^{(k+1)} \right]^{z_{ig}}  , \label{eq:complete-data likelihood2}
\end{align}
where 
\begin{align*}
\phi_d\left(\bx_i|\bu_{ig};\bmu_g^{(k+1)},\bLambda_g,\bPsi_g\right) &= \frac{1}{|2 \pi \bPsi_g|^{1/2}} \exp \left\{ - \frac{1}{2} (\bx_i - \bmu_g^{(k+1)}-\bLambda_g \bu_{ig})' \bPsi_g^{-1} (\bx_i - \bmu_g^{(k+1)}- \bLambda_g \bu_{ig}) \right\}. \\
\phi_q (\bu_{ig}) & = \frac{1}{(2 \pi)^{q/2}} \exp \left\{ - \frac{1}{2}\bu_{ig}' \bu_{ig} \right\}. 
\end{align*}
Now the complete data log-likelihood is given by
\begin{align}
\cL_{c2} (\btheta_2)& =-\frac{nd}{2} \ln 2 \pi +\sum_{g=1}^G n_g \ln \pi_g  +\frac{1}{2}\sum_{i=1}^n \sum_{g=1}^G z_{ig} \ln | \bPsi^{-1}_g| \nonumber \\
 & \quad -\frac{1}{2} \sum_{i=1}^n\sum_{g=1}^G z_{ig}\tr\left\{ (\bx_i - \bmu_g^{(k+1)}- \bLambda_g \bu_{ig}) (\bx_i - \bmu_g^{(k+1)}- \bLambda_g \bu_{ig})' \bPsi_g^{-1} \right\}. \label{L2(theta2)}
\end{align}
Some algebras lead to the following estimate of $\{  (\bLambda_g$, $\bPsi_g), \,  g=1, \ldots, G \}$:
\begin{align*}
\hat{\bLambda}_g  & = \bS^{(k+1)}_g \bgamma^{(k)'}_g [\bTheta_g^{(k)}]^{-1} \\
 \hat{\bPsi}_g & =\text{diag}\left\{ \bS^{(k+1)}_g- \hat{\bLambda}_g \bgamma_g^{(k)} \bS^{(k+1)}_g\right\} \, . 
\end{align*}
where we set
\begin{align*}
\bS_g^{(k+1)} & =(1/n_g^{(k+1)})\sum_{i=1}^n z_{ig}^{(k+1)}(\bx_i - \bmu_g^{(k+1)}) (\bx_i - \bmu_g^{(k+1)})'  \\
\bgamma^{(k)}_g & =\bLambda^{(k)'}_g (\bLambda^{(k)}_g\bLambda^{(k)'}_g+\bPsi^{(k)}_g)^{-1} \\
\bTheta^{(k)}_{ig} & =\bI_q-\bgamma^{(k)}_g \bLambda^{(k)}_g +\bgamma^{(k)}_g (\bx_i-\bmu_g)(\bx_i-\bmu_g)' \bgamma^{(k)'}_g . 
\end{align*}
Hence the maximum likelihood estimates $\hat{\bLambda}_g$  and  $\hat{\bPsi}_g$ for $\bLambda$ and $\bPsi$ can be obtained by alternatively computing the update estimates $\bLambda_g^{+} $ and $\bPsi^{+}_g$, by
\begin{align}
\bLambda_g^{+} & = \bS^{(k+1)}_g \bgamma^{(k)'}_g [\bTheta_g^{(k)}]^{-1} \qquad \mbox{and} \qquad \bPsi^{+}_g  =\text{diag}\left\{ \bS^{(k+1)}_g- \bLambda_g^{+} \bgamma^{(k)}_g\bS^{(k+1)}_g\right\} \, , \label{LambdaAndPsi}
\end{align}
and, from the latter, evaluating the  update estimates $\bgamma_g^{+}$ and $\Theta_g^{+}$ by
\begin{align}
 \bgamma_g^{+} = \bLambda_g^{'} ( \bLambda_g \bLambda_g^{'} +\bPsi_g)^{-1} \qquad \mbox{and} \qquad 
\bTheta_g^{+}=\bI_q-\bgamma_g \bLambda_g+ \bgamma_g \bS_g^{(k+1)} \bgamma^{'}_g, \label{gammaAndTheta}
\end{align}
iterating these two steps until convergence on $\hat{\bLambda}_g$ and $\hat{\bPsi}_g$, so giving ${\bLambda}^{(k+1)}_g$ and ${\bPsi}^{(k+1)}_g$ .

In summary, the procedure can be described as follows. For a given initial random clustering $\bz^{(0)}$, on the $(k+1)-th$ iteration, the algorithm carries out the following steps, for $g=1, \ldots, G$:
\begin{enumerate}
\item Compute $z_{ig}^{(k+1)}$  and consequently obtain $\pi^{(k+1)}_g$, $\bmu^{(k+1)}_g$, $n_g^{(k+1)}$ and $\bS^{(k+1)}_g$;
\item Set a starting value for $\bLambda_g $ and $\bPsi_g $ from $\bS^{(k+1)}_g$;
\item Repeat the following steps,  until convergence on  $\hat{\bLambda}_g$ and $\hat{\bPsi}_g$:
\begin{enumerate}
\item Compute $\bgamma_g^{+}$ and $\bTheta_g^{+}$ from (\ref{gammaAndTheta});
\item Set $\bgamma_g \leftarrow \bgamma^{+}_g$ and $\bTheta_g \leftarrow \bTheta^{+}_g$;
\item Compute 
$\bLambda_g^{+} \leftarrow  \bS^{(k+1)}_g \bgamma^{'}_g (\bTheta_g^{-1})$ and $\bPsi^{+}_g  \leftarrow \text{diag}\left\{ \bS^{(k+1)}_g- \bLambda_g^{+} \bgamma^{}_g \bS^{(k+1)}_g\right\}$;
\item Set $\bLambda_g \leftarrow \bLambda_g^{+}$ and $\bPsi_g \leftarrow \bPsi_g^+$;
\end{enumerate}
\end{enumerate}

To completely describe the algorithm, here we give more details on how  to specify  the starting values for $\bLambda_g $ and $\bPsi_g $ from $\bS^{(k+1)}_g$, as it is needed in Step 2.

Starting from  the eigen-decomposition of $\bS^{(k+1)}_g$, say $\bS^{(k+1)}_g=\bA_g \bB_g \bA_g'$, computed on the base of  $z^{(k+1)}_{ig}$, the main idea is that $\bLambda_g$ has to synthesize the "more important" relations between the $d$ observed features, see \cite{McNi:Murp:Pars:2008}. Then, looking at the equality $\bSigma_g=\bGamma_g\bGamma_g'+\bPsi_g$, the initial values of $\bLambda_g$ were set as 
\begin{equation}
\lambda_{ij}=\sqrt{d_j}a_{ij}
\end{equation} 
where $d_{j}$ is the $j$th largest eigenvalue of $\bS^{(k+1)}_g$ and $a_{ij}$ is the $i$th element of the corresponding eigenvector $\ba_j$ (the $j$th column in $A_g$), for $i \in\{1,2,\ldots,p\}$ and $j \in\{1,2,\ldots,q\}$. Finally the $\bPsi_g$ matrices can be initialized by the position  $\bPsi_g=\diag \{\bS^{(k+1)}_g-\bLambda_g\bLambda_g'\}$.

\section{Likelihood maximization in constrained parametric spaces} \label{ConstrainedML}

Properties of maximum likelihood estimation for normal mixture models have been deeply investigated. It is well known that $\cL(\theta)$ is unbounded on $\bTheta$ and may present many local maxima. Day (1969) was perhaps the first noting that any small number of sample points, grouped sufficiently close together, can give raise to spurious maximizers, corresponding to parameters points with greatly differing component standard deviation.
To overcome this issue and to prevent $\cL(\theta)$ from singularities, \citet{Hath:Acon:1985} proposed a constrained maximum likelihood formulation for mixtures of univariate normal distributions, suggesting a natural extension to the multivariate case. 
Let $c \in (0,1]$, then the following constraints
\begin{equation}
\min_{1 \leq h \neq j \leq k} \lambda (\bSigma_h \bSigma_j^{-1}) \geq c   \label{constlambda}
\end{equation}
on the eigenvalues $\lambda$ of $\bSigma_h \bSigma_j^{-1}$ leads to properly defined, scale-equivariant, consistent ML-estimators for the mixture-of-normal case, see Hennig (2004). 
It is easy to show  that a sufficient condition for (\ref{constlambda}) is
\begin{equation}
a \leq \lambda_{ig} \leq b  , \qquad i=1,\ldots,d; \qquad g= 1,\ldots,G \label{alambdab}
\end{equation}
where $\lambda_{ig}$ denotes the $i$th eigenvalue of $\bSigma_g$ i.e. $\lambda_{ig}=\lambda_i(\bSigma_g)$, and for $a,b \in \mR^{+}$ such that $a/b \geq c$, see \citet{Ingr:2004}. Differently from (\ref{constlambda}),  condition (\ref{alambdab}) can  be easily implemented in any optimization algorithm.
Let us consider the constrained parameter space $\bTheta_c$ of $\bTheta$:
\begin{align}
\bTheta_c =&  \{ (\pi_1, \ldots, \pi_G,  \bmu_1, \ldots, \bmu_G, \bSigma_1, \ldots, \bSigma_G) 
\in \mathbb{R}^{k[1+d+(d^2+d)/2]} \: : \:  \nonumber
\\& \pi_g \geq 0,  \: \pi_1+\cdots+\pi_G =1, 
\: a \leq \lambda_{ig} \leq b, \quad  g=1, \ldots, G \; \; \; i=1, \ldots, d\}.  \label{Psi_c} 
\end{align}
Due to the structure of the covariance matrix $\bSigma_g$ given in \eqref{Sigmag},  bound in  \eqref{alambdab} yields
\begin{equation}
 \lambda_{\rm min} (\bLambda_g \bLambda'_g+\bPsi_g) \geq a \qquad \mbox{and} \qquad \lambda_{\rm max} (\bLambda_g \bLambda'_g+\bPsi_g) \leq b  , \qquad g=1,\ldots,G \label{alambdabSigma}
\end{equation}
where $\lambda_{\rm min} (\cdot)$ and $\lambda_{\rm max} (\cdot)$ denote the smallest and the largest eigenvalue of $(\cdot)$ respectively. 
Since $\bLambda_g \bLambda'_g$ and $\bPsi_g$  are  symmetric and positive definite, then it results:
\begin{align}
\lambda_{\rm  min} (\bLambda_g \bLambda'_g+\bPsi_g) & \geq  \lambda_{\rm  min} (\bLambda_g \bLambda'_g) + \lambda_{\rm min} (\bPsi_g)  \geq a \label{lminSigma}\\
\lambda_{\rm  max} (\bLambda_g \bLambda'_g+\bPsi_g) & \leq  \lambda_{\rm  max} (\bLambda_g \bLambda'_g)  + \lambda_{\rm max} (\bPsi_g)  \leq b \, , \label{lmaxSigma}
\end{align}
see \citet{Lutk:Matrix:1996}. 
 Moreover, being  $\bPsi_g$ a diagonal matrix, then
\begin{alignat}{2}
\lambda_{\rm min} (\bPsi_g) & = \min_i \psi_{ig} & \qquad \mbox{and} \qquad  
\lambda_{\rm max} (\bPsi_g) &= \max_i \psi_{ig}, \label{lminmaxPsi}
\end{alignat}
where $\psi_{ig}$ denotes the $i$-th diagonal entry of the matrix $\bPsi_{g}$.

Concerning the square $d \times d$ matrix $\bLambda_g \bLambda'_g$ ($g=1,\ldots, G$), we can get its eigenvalue  decomposition, 
i.e. we can find $\bLambda_g$ and $\bGamma_g$ such that
\begin{equation}
\bLambda_g \bLambda'_g =\bGamma_g \bDelta_g \bGamma'_g  \label{decompSigma1}
\end{equation} 
where $\bGamma_g$ is the orthonormal matrix whose rows are the eigenvectors of $\bLambda_g \bLambda'_g$ and $\bDelta_g=\diag(\delta_{1g}, \ldots, 
\delta_{dg})$ is the diagonal matrix of the eigenvalues of $\bLambda_g \bLambda'_g$, sorted in non increasing order, i.e. $\delta_{1g}\geq \delta_{2g} \geq \ldots \geq \delta_{qg} \geq 0$,
and $\delta_{(q+1)g} = \cdots = \delta_{dg}=0$.  

Now, we can apply the singular value decomposition to the $d \times q$ rectangular matrix $\bLambda_g$, so giving $\bLambda_g = \bU_g \bD_g \bV'_g$, where $\bU_g$ is a $d \times d$ unitary matrix (i.e., such that $\bU'_g \bU_g = \bI_d$) and $\bD_g$ is a $d \times q$ rectangular diagonal matrix with $q$ nonnegative real numbers on the diagonal, known as \textit{singular values}, and $\bV_g$ is a $q \times q$ unitary matrix. The $d$ columns of $\bU$ and the $q$ columns of $\bV$ are called the \textit{left singular vectors} and \textit{right singular vectors} of $\bLambda_g$, respectively. Now we have that 
\begin{equation}
\bLambda_g \bLambda'_g = (\bU_g \bD_g \bV'_g)(\bV_g \bD'_g \bU'_g) = \bU_g \bD_g \bI_q \bD'_g \bU'_g =  \bU_g \bD_g  \bD'_g \bU'_g \label{decompSigma2}
\end{equation}
and equating \eqref{decompSigma1} and \eqref{decompSigma2} we get $\bGamma_g= \bU_g$ and $\bDelta_g = \bD_g\bD'_g $, that is 
\begin{equation}
\diag(\delta_{1g}, \ldots, \delta_{qg}) = \diag(d_{1g}^2, \ldots, d_{qg}^2) \, . \label{eqeigen}
\end{equation} 
with $d_{1g} \geq d_{2g} \geq \cdots \geq d_{qg} \geq 0$.  
In particular, it is known that only the first $q$ values of $\bD_g$ are non negative, and the remaining $d-q$ terms are null.
Thus it results
\begin{equation}
\lambda_{\rm  max} (\bLambda_g \bLambda'_g)  = d_{1g}^2 .\label{lmaxLambda}
\end{equation}

Supposing now to choose a value for the upper bound $b$ in such a way that ${b \geq \bPsi_{ig}}$ for $g=1, \ldots, G$ and $i=1, \ldots, q$, then  constraints  
\eqref{lminSigma} and \eqref{lmaxSigma} are satisfied when 
\begin{align} 
d_{ig}^2+ \psi_{ig} & \geq a \quad \quad \quad \quad &i=1,\ldots, d \label{lminLambdaPsia}    \\
d_{ig} & \leq \sqrt{b - \bPsi_{ig}}  & i=1,\ldots,q \label{lmaxPsib} \\
\psi_{ig} & \leq  b \quad & i=q+1,\ldots,d \label{lmaxPsi} 
\end{align}
for $g=1, \ldots, G$.  
In particular, we remark that  condition  \eqref{lminLambdaPsia} reduces to $\Psi_{ig}\geq a$ for $i=(q+1), \ldots, d$.

\section{Constraints on the covariance matrix for factor analyzers}\label{sec:constraintsFA}

The two-fold (eigenvalue and singular value) decomposition of the $\bLambda_g$ presented above, suggests how to  modify the EM algorithm in such a way that the eigenvalues  of the covariances $\bSigma_g$ (for $g=1,\dots,G$) are confined into suitable ranges. To this aim we have to implement  constraints  \eqref{lminLambdaPsia},  \eqref{lmaxPsib}  and \eqref{lmaxPsi}. 

We proceed  as follows on the $(k+1)$th iteration:
\begin{enumerate}
\item Decompose $\bLambda_g$ according to the singular value decomposition as $\bLambda_g = \bU_g \bD_g \bV'_g$;
\item Compute the squared singular values  $(d_{1g}^2, \ldots, d_{qg}^2)$ of $\bLambda_g$;
\item Create a copy $\bD^*_g$ of $\bD_g^{(k+1)}$ and a copy $\bPsi^*_g$ of $\bPsi^{(k+1)}_g$;
\item  For $i=1$ to $q$, if $d_{ig}^2 + \psi_{ig}^{(k+1)} < a$, then if $a-\psi_{ig}^{(k+1)} \geq 0$ set  $d_{ig} \leftarrow \sqrt{a-\psi_{ig}^{(k+1)}}$  else $d_{ig} \leftarrow \sqrt{a}$  into $\bD^*_g$; 
\item For $i=q+1$ to $d$, if $\psi_{ig}^{(k+1)} <  a$ then set $\psi_{ig}^{(k+1)} \leftarrow  a$ into $\bPsi^*_g$; 
\item  For $i=1$ to $q$, if $d_{ig}^2+\psi_{ig}^{(k+1)} > b $, then if $b-\psi_{ig}^{(k+1)} \geq 0$ set  $d_{ig} \leftarrow \sqrt{b - \psi_{ig}^{(k+1)} }$  into $\bD^*_g$ else $d_{ig} \leftarrow \sqrt{b}$  into $\bD^*_g$; 
\item For $i=q+1$ to $d$, if $\psi_{ig}^{(k+1)} >  b$ then set $\psi_{ig}^{(k+1)} \leftarrow  b$ into $\bPsi^*_g$; 
\item  Set $\bLambda^{(k+1)}_g \leftarrow \bU_g \bD^*_g \bV'_g$;
\item Set $\bPsi^{(k+1)}_g \leftarrow \bPsi^*_g$. 
\item Stop.
\end{enumerate}
It is important to remark that the resulting EM algorithm is monotone, once the initial guess, say $\bSigma_g^0$, satisfies the constraints. Further, as shown in the case of gaussian mixtures in \cite{Ingr:Rocc:2007}, the maximization of the complete loglikelihood is guaranteed. From the other side, it is apparent that the above recipes require some a priori information on the covariance structure of the mixture, throughout the bounds $a$ and $b$.  

\section{Numerical studies}\label{sec:numerical results}

In this section we present numerical studies, based on both simulated and real data sets, in order to show the performance of the constrained EM algorithm with respect to
unconstrained approaches. 

\subsection{Artificial data}\label{sec:simdata}
We  consider here three mixtures of $G$ 
components of $d$-variate normal distributions, for different values of the parameter   $\btheta_0$. 
First, we point out that the point of local maximum corresponding to the consistent estimator $\btheta^*$, has been chosen to be the limit
of the EM algorithm using the true parameter $\btheta_0$ as initial estimate, i.e. considering the true classification. In other words, we set  $z_{ig}=1$ if the $i$th unit comes from the $g$th component and  $z_{ig}=0$ otherwise. In the following, such estimate will be referred to as the right maximum of the likelihood function.

To begin with, we  generate a set of 100 different random initial clusterings to initialize the algorithm at each run. To this aim, for a fixed
number $G$ of components of the mixture, we draw each time a set of random starting values for the $z_{ig}$
from the multinomial distribution with values in $(1, 2, \ldots, G)$ with parameters $(p_1, p_2, \ldots, p_g) = (1/G, 1/G,\ldots, 1/G)$.  Then we run a hundred times  both the unconstrained  and the
constrained AECM algorithms (for different values of the constraints $a,b$)  using the same set of initial clusterings in both cases. 
The initial values for the elements of $\bLambda_g$ and $\bPsi_g$  can be obtained as described at the end of Section \ref{sec:The AECM algorithm} from the eigen-decomposition of $\bS_g$, and the algorithms run until convergence or it reaches  the fixed maximum number of iterations. 

The stopping criterion is based on the Aitken acceleration procedure \citep{Aitk:OnBe:1926}, to estimate the asymptotic maximum of the log-likelihood at each iteration of the EM algorithm
(in such a way,  a decision can be made regarding whether or not the algorithm reaches convergence;
that is, whether or not the log-likelihood is sufficiently close to its estimated asymptotic value). 
The Aitken acceleration at iteration $k$ is given by
\begin{displaymath}
	a^{\left(k\right)}=\frac{\cL^{\left(k+1\right)}-\cL^{\left(k\right)}}{\cL^{\left(k\right)}-\cL^{\left(k-1\right)}},
\end{displaymath}
where $\cL^{\left(k+1\right)}$, $\cL^{\left(k\right)}$, and $\cL^{\left(k-1\right)}$ are the log-likelihood values from iterations $k+1$, $k$, and $k-1$, respectively. 
Then, the asymptotic estimate of the log-likelihood at iteration $k + 1$ is given by
\begin{displaymath}	\cL_{\infty}^{\left(k+1\right)}=\cL^{\left(k\right)}+\frac{1}{1-a^{\left(k\right)}}\left(\cL^{\left(k+1\right)}-\cL^{\left(k\right)}\right),
\end{displaymath}
see \citet{Bohn:Diet:Scha:Schl:Lind:TheD:1994}.
In our analyses, the algorithms stop when $\cL_{\infty}^{\left(k+1\right)}-\cL^{\left(k\right)}<\epsilon$, with $\epsilon=0.001$.
Programs have been written in the R language; the different cases and the obtained results are described below.

\paragraph{\textsc{Mixture 1: $G=3$, $d=6$, $q=2$, $N=150$.}} \ \\

The sample has been generated with weights $\balpha = (0.3, 0.4, 0.3)'$ according to the following parameters:

  \begin{align*}
 \bmu_1 &= (0,0,0,0,0,0)'  &  \bPsi_1 &= \mbox{diag}(0.1,0.1,0.1,0.1,0.1,0.1) \\
 \bmu_2 &= (5,5,5,5,5,5)'  & \bPsi_2 &= \mbox{diag}(0.4,0.4,0.4,0.4,0.4,0.4) \\
 \bmu_3 &= (10,10,10,10,10,10)'  & \bPsi_3 &= \mbox{diag}(0.2,0.2,0.2,0.2,0.2,0.2) 
  \end{align*}
  \vspace{-13mm}\\
 \begin{gather*}
\bLambda_1 = \begin{pmatrix} 0.50 & 1.00 \\ 1.00 & 0.45 \\  0.05 & -0.50 \\ -0.60 & 0.50 \\ 0.50 & 0.10 \\ 1.00 & -0.15 \end{pmatrix}  \quad  \quad
\bLambda_2 = \begin{pmatrix} 0.10 & 0.20 \\ 0.20 & 0.50 \\  1.00 & -1.00 \\ -0.20 & 0.50 \\ 1.00 & 0.70 \\ 1.20 & -0.30 \end{pmatrix} \quad  \quad
 \bLambda_3 = \begin{pmatrix} 0.10 & 0.20 \\ 0.20 & 0.00 \\  1.00 & 0.00 \\ -0.20 & 0.00 \\ 1.00 & 0.00 \\ 0.00 & -1.30 \end{pmatrix}.\\
\end{gather*}

  \vspace{-5mm}
Hence, the covariance matrices  $\bSigma_g = \bLambda_g \bLambda'_g + \bPsi_g$ ($g=1, 2, 3$)  have  the following eigenvalues:
\begin{align*}
\lambda(\bSigma_1) &=(3.17, 1.63, 0.10, 0.10, 0.10, 0.10)' \\
\lambda(\bSigma_2) &=(4.18, 2.27,0.40, 0.40, 0.40,0.40)' \\
\lambda(\bSigma_3) &=(2.29, 1.93, 0.20,0.20, 0.20, 0.20)',
\end{align*}
whose  largest value  is given by $\max_{i,g} \lambda_i(\bSigma_g)=4.18 \, . $

\begin{figure} 
	\begin{center}
		\includegraphics[width=9 cm, height=8 cm]{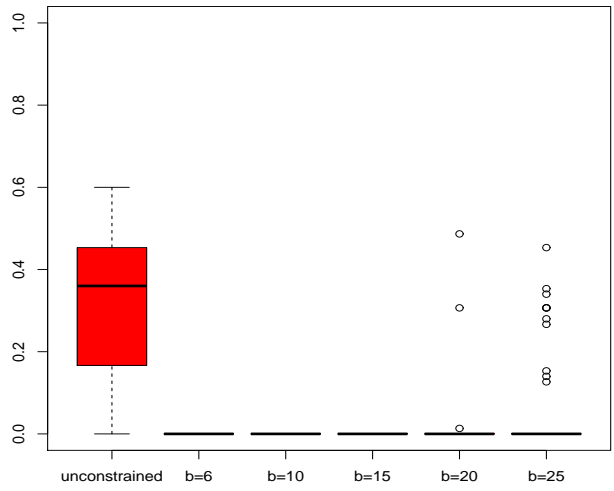} 
			\end{center}
	\caption{\rm Mixture 1: Boxplots of the misclassification error. From left to right, the first boxplot refers to the unconstrained algorithm, then the following boxplots correspond to the constrained algorithm, for $a=0.01$ and $b$ respectively set to the values  $b=6,10,15,20,25$.}\label{fig:BoxPlotMixture1}
\end{figure}

First we run the unconstrained algorithm: the right solution has been attained in 24\% of cases, without incurring in  singularities. Summary statistics (minimum, first quartile $Q_1$, median $Q_2$, third quartile
$Q_3$ and maximum) about  the distribution of the misclassification error over the 100 runs are reported in Table \ref{tab:MisclassCase1}.
Due to the choice on parameters, we rarely expect  too small eigenvalues in the estimated covariance matrices: we set $a=0.01$ to protect from them; conversely, as local maxima are quite often due to  large estimated eigenvalues, we consider setting also a constraint from above, taking into account some values for $b$, the upper bound. To compare how the choice of the bounds  $a$ and $b$  influences the performance of the constrained EM, we experimented with different pairs of values, and in Table \ref{tab:case1} we report the more interesting cases.  Further results are reported in Figure \ref{fig:BoxPlotMixture1}, which provides the boxplots of the distribution of the misclassification errors obtained in the  sequence of $100$ runs,
showing the  poor performance of the unconstrained algorithm compared with the good behaviour of its  constrained version. For all values of the upper bound $b$, the third quartile of the misclassification error is steadily equal to $0$.  Indeed, for $b=6,10$ and 15 we  had no misclassification error,  while we observed very low and rare misclassification errors only for $b=20$ and $b=25$ (respectively 3 and 11 not null values, over 100 runs). Moreover, the robustness of the results with respect to the choice of the upper constraint is apparent.
\begin{table}[h]
\begin{center}
\caption{Mixture 1: Summary  statistics of the distribution of the Misclassification Error over 100 runs of the unconstrained EM algorithm}\label{tab:MisclassCase1}
\begin{tabular}{ccccc} \hline
  \multicolumn{5}{c}{Misclassification Error}  \\
  min &  $Q_1$& $Q_2$ & $Q_3$ & max \\ \hline
  0\% & 17\% & 36\% & 45.3\% & 60\% \\
  \hline\hline
\end{tabular}
\end{center}
\end{table}
\begin{table}[h]
\begin{center}
\caption{Mixture 1: Percentage of convergence to the right maximum of the constrained EM algorithms
for $a=0.01$ and some values of the upper constraint $b$}\label{tab:case1}
\begin{tabular}{cccccc} \hline
  \multicolumn{6}{c}{$b$}  \\
$+\infty$ &  6 &  10 & 15 & 20 & 25 \\ \hline
24\% &  100\% & 100\% & 100\% & 97\% & 89\% \\
  \hline\hline
\end{tabular}
\end{center}
\end{table}

\begin{figure} 
	\begin{center}
		\includegraphics[width=0.32\textwidth] {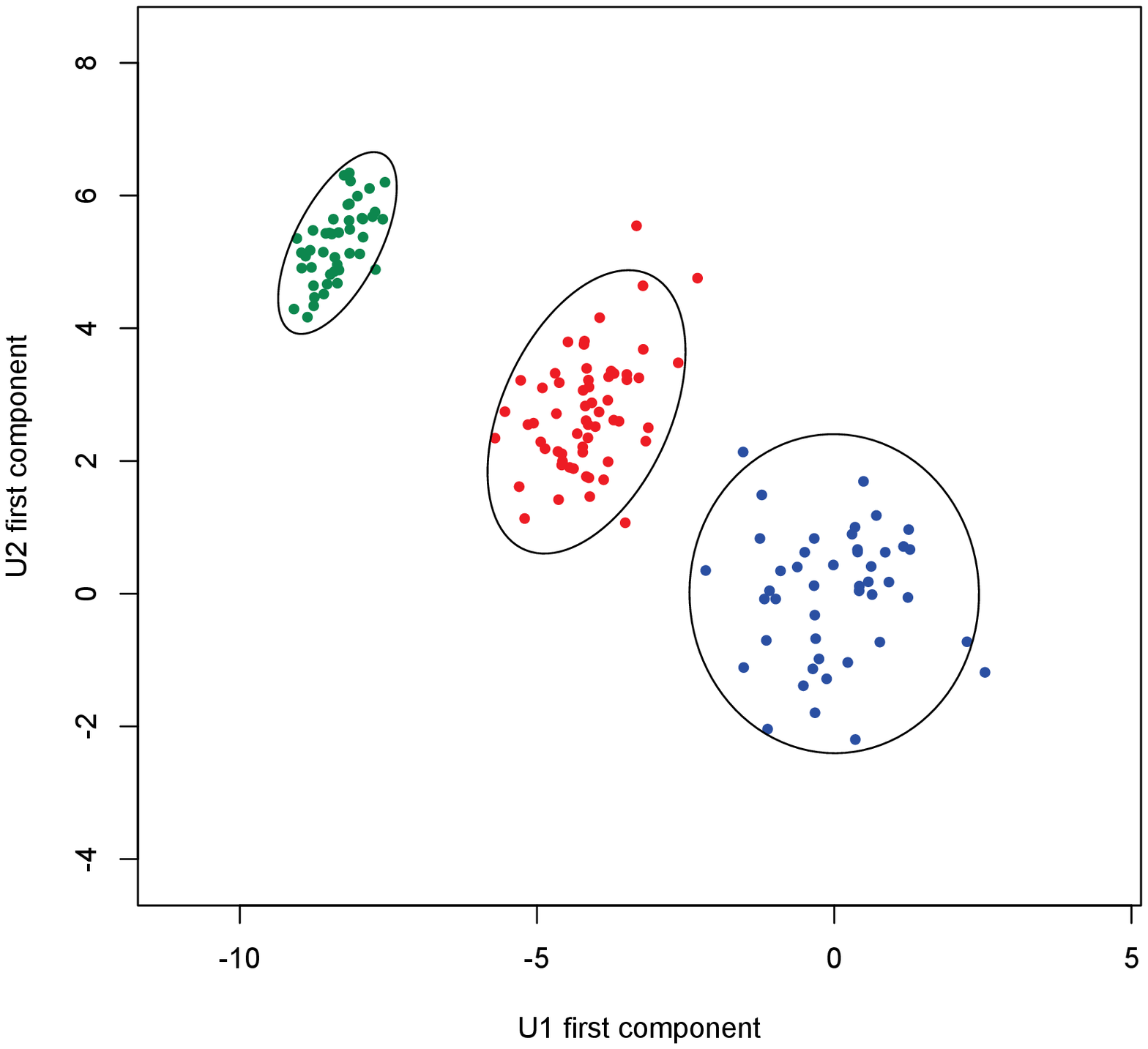} 
		\includegraphics[width=0.32\textwidth] {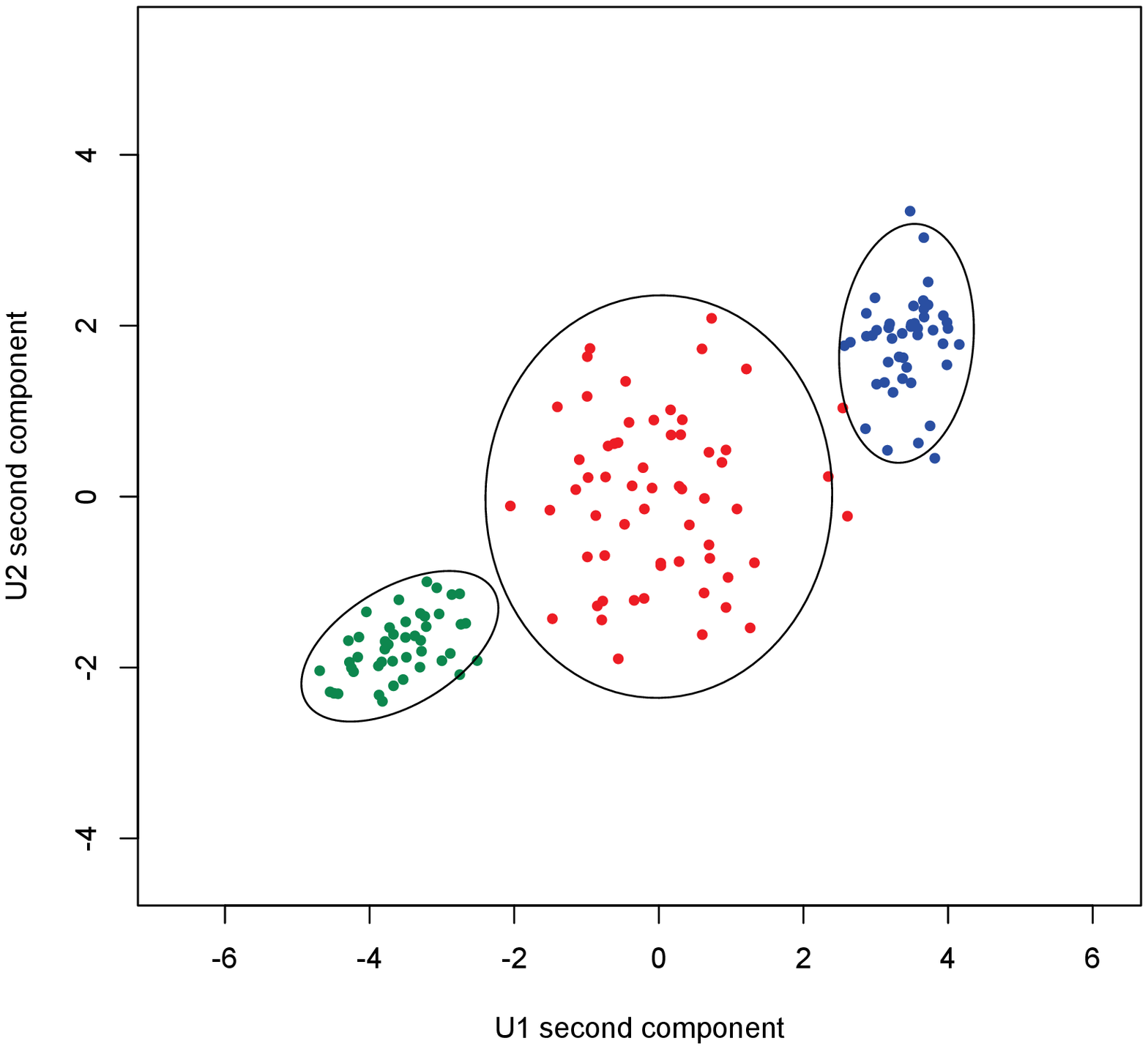} 
		\includegraphics[width=0.32\textwidth] {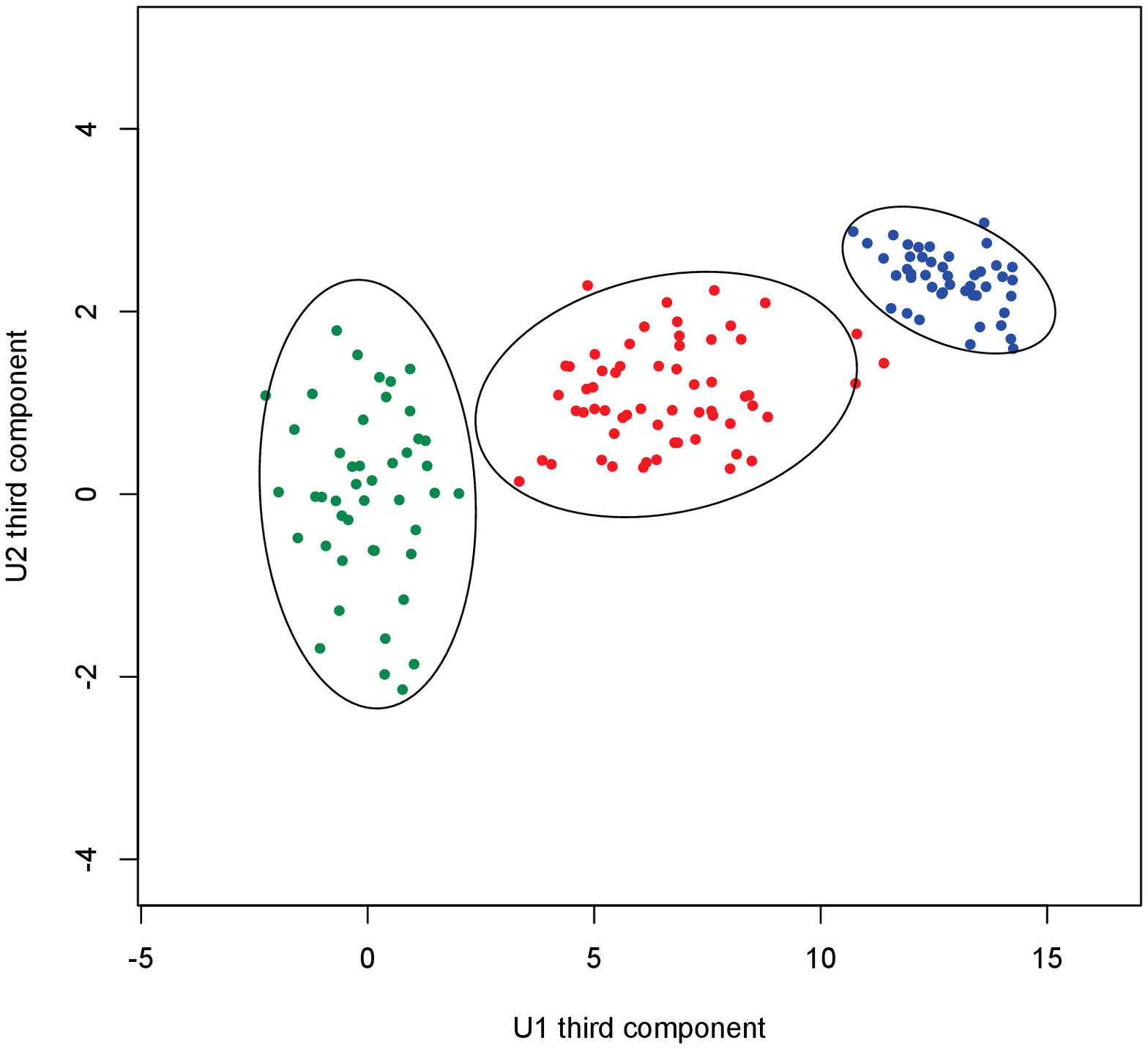} \\
		\includegraphics[width=0.32\textwidth] {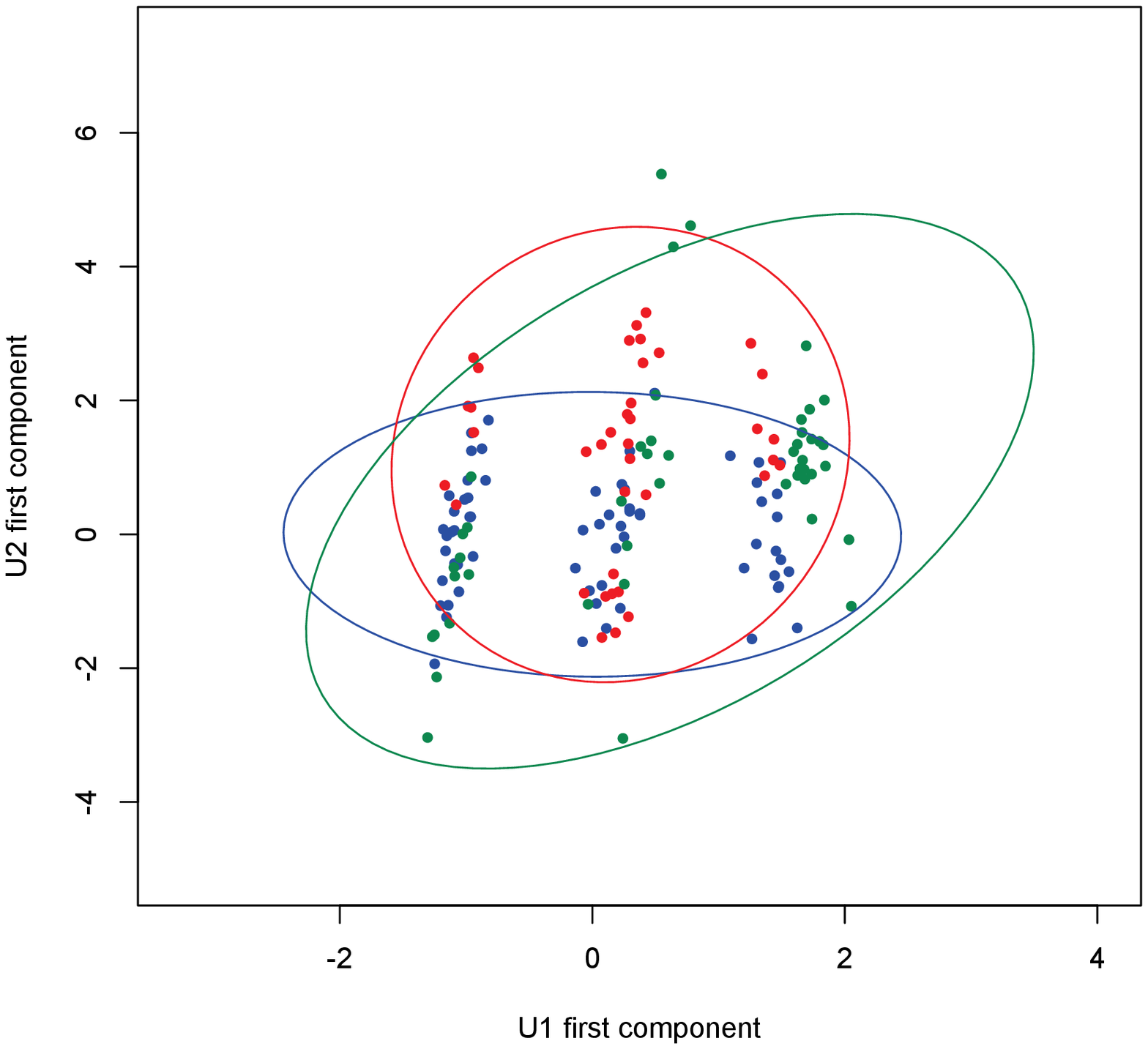}
		 \includegraphics[width=0.32\textwidth] {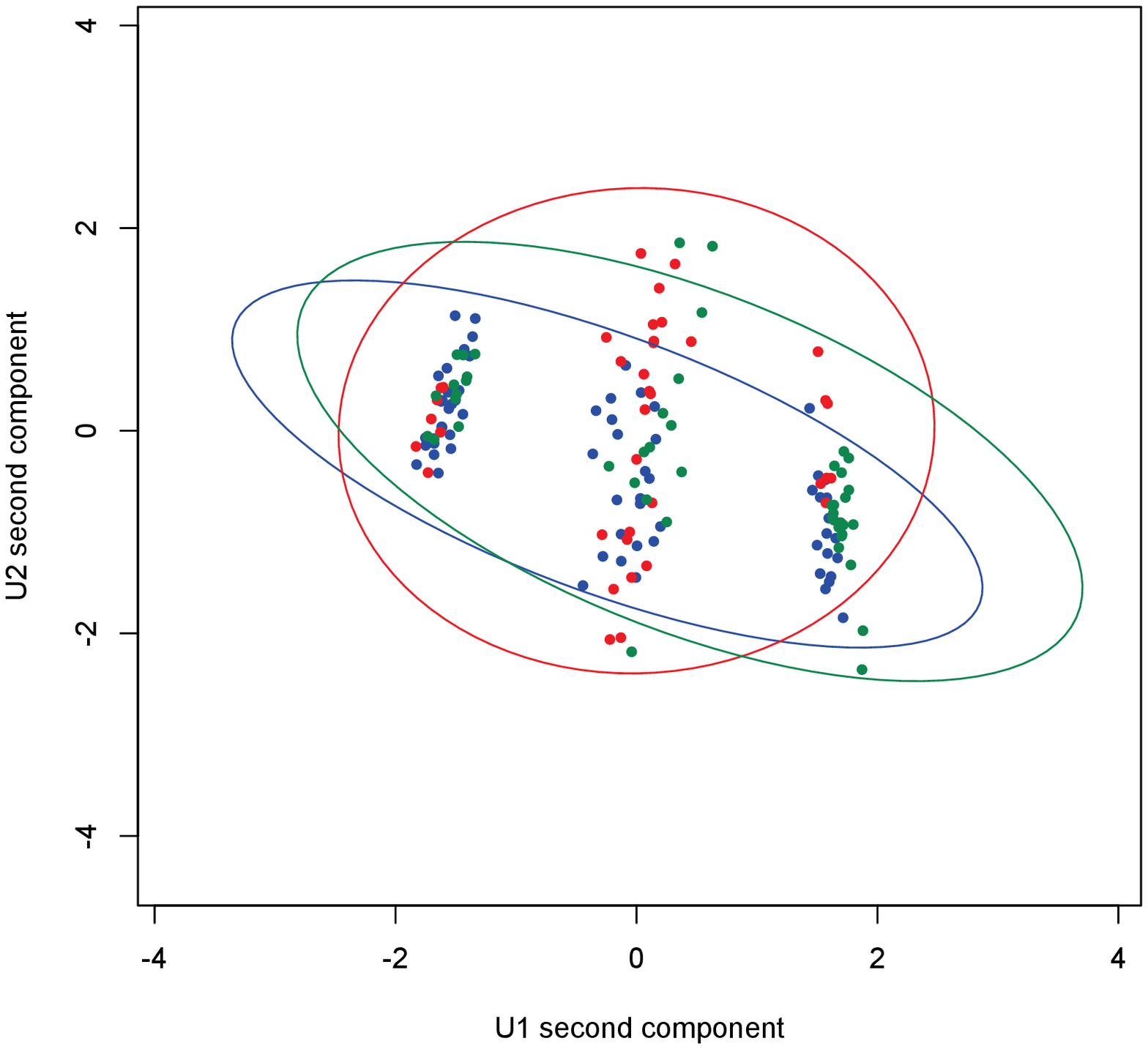} 
		 \includegraphics[width=0.32\textwidth] {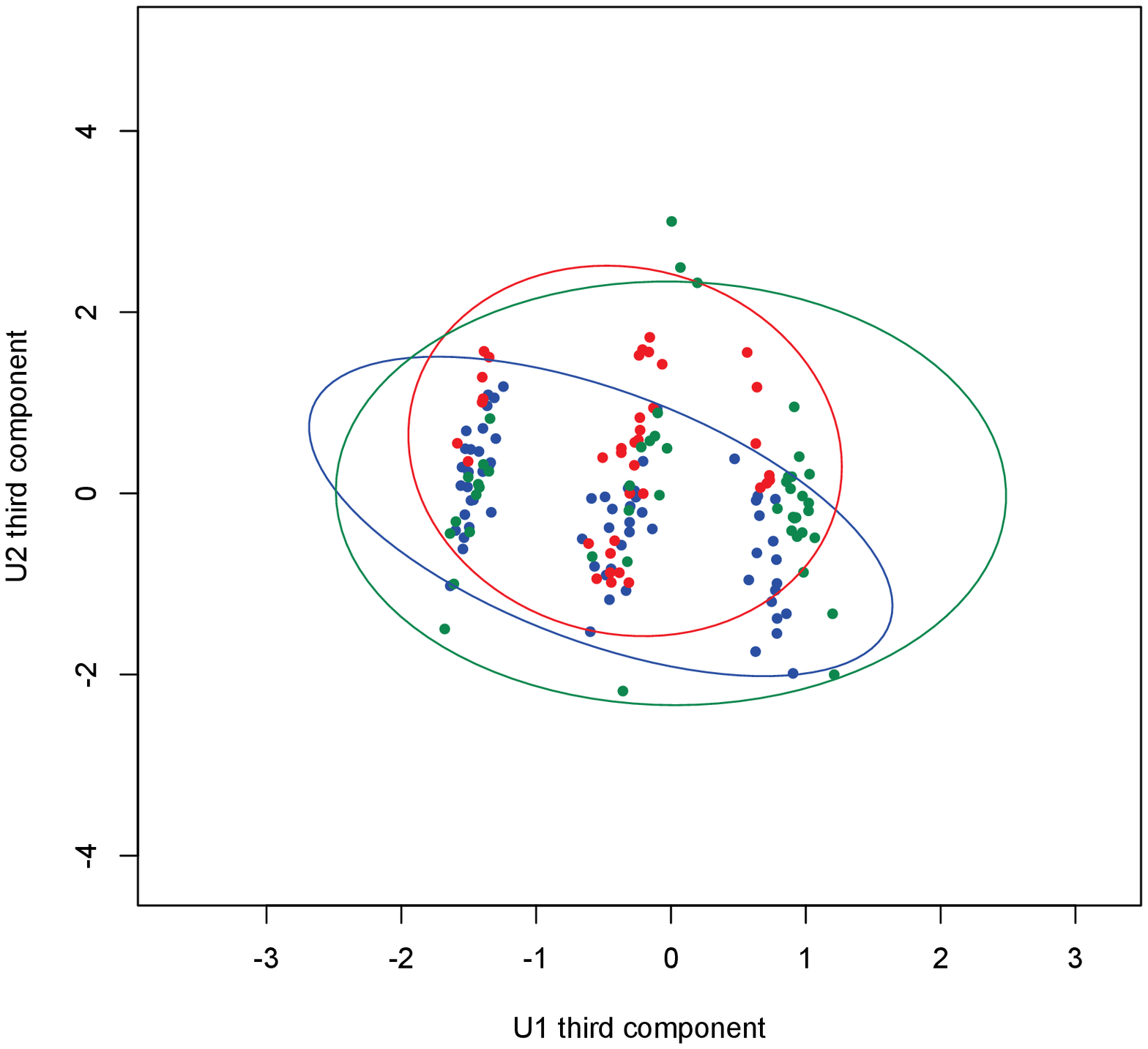}
	\end{center}
	\caption{\rm Mixture 1: plot of the classified data on the three factor spaces, under the true maximum of the likelihood function (upper row) and, conversely, under  a spurious maximum of the likelihood function (row below)}
	\label{fig:Mixture_1A}
\end{figure}

In Figure \ref{fig:Mixture_1A} we plot the classified data on the three factor spaces given by $\hat{\bU}_{i1}, \hat{\bU}_{i2}$ and $\hat{\bU}_{i3}$
under  the true maximum of the likelihood function (first rows of plots), while in the second row we give the classification obtained
according to a spurious maximum of the likelihood function.

We recall that an original data point $\bx_i$ can be represented in $q$ dimensions by the posterior distribution of its associated $q$-dimensional latent factor $\bU_i$. A convenient summary of this distribution is its mean. Hence we can portray the $\bx_i$ in $q$-dimensional space by plotting the estimated conditional expectation of each $\bU_i$ given $\bx_i$, that is, the (estimated) posterior mean of the factor $\bU_i$ (for $i=1,\ldots,n$). We have that 
\begin{align*}
\hat {\bu_i} &= \EE_{\hat{\btheta}} \{\bU_i \| \bx_i \} =\hat{\gamma(\bx_i-\overline{\bx}})
\end{align*}
where $\EE_{\hat{\btheta}} $ denotes expectation using the estimate $\hat{\btheta}$ instead of $\btheta$, and  $\hat{\gamma}$ has been computed following (\ref{gammaAndTheta}).\\
In the particular case of $q=2$, as in this simulation experiment, we can draw the data in a bidimensional plot  in Figure \ref{fig:Mixture_1A}. From the two series of plots, it can be seen that the appropriate factor space allows for the right classification, while a spurious likelihood maximizer leads to unsuitable factor spaces, which in turn generate serious issues in classification. 

\paragraph{\textsc{Mixture 2: $G=4$, $d=7$, $q=2$, $N=100$.}}\ \\

The sample has been generated with weights $\balpha = (0.2,0.3,0.35,0.15)'$ according to the following parameters:
 \begin{align*}
\bmu_1 &= (0,0,0,0,0,0,0)'  & \bPsi_1 &= \mbox{diag}(0.2,0.2,0.2,0.2,0.2,0.2,0.2)\\ 
 \bmu_2 &= (5,5,5,5,5,5,5)'  & \bPsi_2 &= \mbox{diag}(0.25,0.25,0.25,0.25,0.25,0.25,0.25) \\ 
\bmu_3 &= (10,10,10,10,10,10,10,)'  & \bPsi_3 &= \mbox{diag}(0.15,0.15,0.15,0.15,0.15,0.15,0.15) \\ 
\bmu_4 &= (15,15,15,15,15,15,15)'  & \bPsi_4 &= \mbox{diag}(0.1,0.1,0.1,0.1,0.1,0.1,0.1) 
\end{align*}
\vspace{-6mm}
\begin{gather*}
\small
\bLambda_1 = \begin{pmatrix} 0.30 & 0.60 \\ 0.60 & 0.27 \\  0.03 & -0.30 \\ -0.36 & 0.30 \\ 0.30&  0.06 \\ 0.60 &-0.09 \\ -0.63 & 1.50 \end{pmatrix}  \,  
\bLambda_2 = \begin{pmatrix} 0.08  &0.16 \\ 0.16 & 0.40 \\   0.80 & -0.80\\ -0.16 & 0.40 \\ 0.80 & 0.56 \\ 0.96 &-0.24\\ 1.60 & -0.24\end{pmatrix} \,
\bLambda_3 = \begin{pmatrix} 0.07 & 0.14\\ 0.14 & 0.00 \\  0.70 & 0.00 \\ -0.14 & 0.00 \\ 0.70 & 0.00 \\ 0.00 & -0.91\\  0.70 & -0.70 \end{pmatrix}\,
\bLambda_4 = \begin{pmatrix} 0.04 & 0.08\\ 0.08 & 0.00 \\  0.40 & 0.00 \\ -0.08 & 0.00 \\ 0.40 & 0.00 \\ 0.00 & -0.52\\  -0.40 & 0.80   \end{pmatrix}.
\normalsize
\end{gather*}
The covariance matrices  $\bSigma_g = \bLambda_g \bLambda'_g + \bPsi_g$ ($g=1, 2, 3$)  have respectively the following eigenvalues:
\begin{align*}
 \lambda(\bSigma_1) &=(4.10	,1.14	,0.33,0.21,	0.15,	0.09,	0.04)'\\
\lambda(\bSigma_2) &=(7.62,	1.18,	0.34,	0.20,	0.18	,0.12,	0.05)' \\
\lambda(\bSigma_3) &=(3.36,	1.36,	0.24,	0.17,	0.14,	0.10,	0.09)' \\
\lambda(\bSigma_4) &=(2.08,	0.48,	0.11,	0.09,	0.07,	0.06,	0.02)'.
\end{align*}
whose  largest value  is given by $\max_{i,g} \lambda_i(\bSigma_g)=7.62 \,.$

\begin{figure} 
	\begin{center}
		\includegraphics[width=9 cm, height=8 cm]{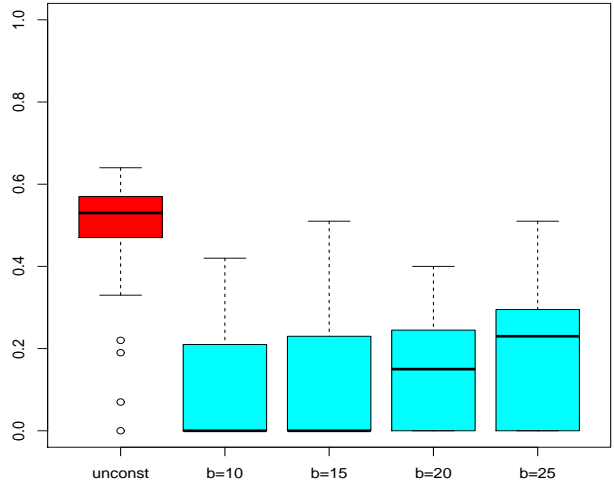} 
			\end{center}
	\caption{\rm Mixture 2: Boxplots of the misclassification error: from left to right, the first boxplot refers to the unconstrained algorithm, then the following boxplots correspond to the constrained algorithm, for $a=0.01$ and $b$ respectively set to the values  $b=10,15,20,25$.}\label{fig:BoxPlotMixture2}
\end{figure}

First we run the unconstrained algorithm: the right solution has been attained only once, over 100 runs. Afterwards, we run the constrained algorithm for different values of the upper bound $b$ on the largest eigenvalue, while maintaining $a=0.01$, and using the same random starting values as before, to compare how the choice of the bounds  influences the performance of the constrained EM. In  Table \ref{tab:case2} we collected the percentage of times in which the algorithm attained the right maximum (where $b=+\infty$ denotes the unconstrained procedure), showing a great improvement with respect to the previous 1\% obtained through the unconstrained version.  
\begin{table}[h]
\begin{center}
\caption{Mixture 2: Percentage of convergence to the right maximum of the constrained EM algorithms
for   $a=0.01$ and different values for the upper bound $b$.}\label{tab:case2}
\begin{tabular}{ccccc} \hline
  \multicolumn{5}{c}{$b$}  \\
  $+\infty$ &10 & 15 & 20 & 25 \\ \hline
  1\%&69\% & 60\% & 46\% & 33\% \\
  \hline\hline
\end{tabular}
\end{center}
\end{table}
Further details are given in Figure \ref{fig:BoxPlotMixture2} which shows the boxplots of the distribution of the  misclassification error in the $5$ sequences of $100$ runs, corresponding to the different values of the constraint $b$. Also in this case the unconstrained algorithm had a bad performance, with a median misclassification error of 0.53, while its constrained version, for $b=10$ and 15, in more than 50\% of the runs had no misclassification error. Furthermore, the unconstrained algorithm did not attain convergence in 4 out of the 100 runs.

\begin{figure} 
	\begin{center}
		\includegraphics[width=0.48\textwidth]{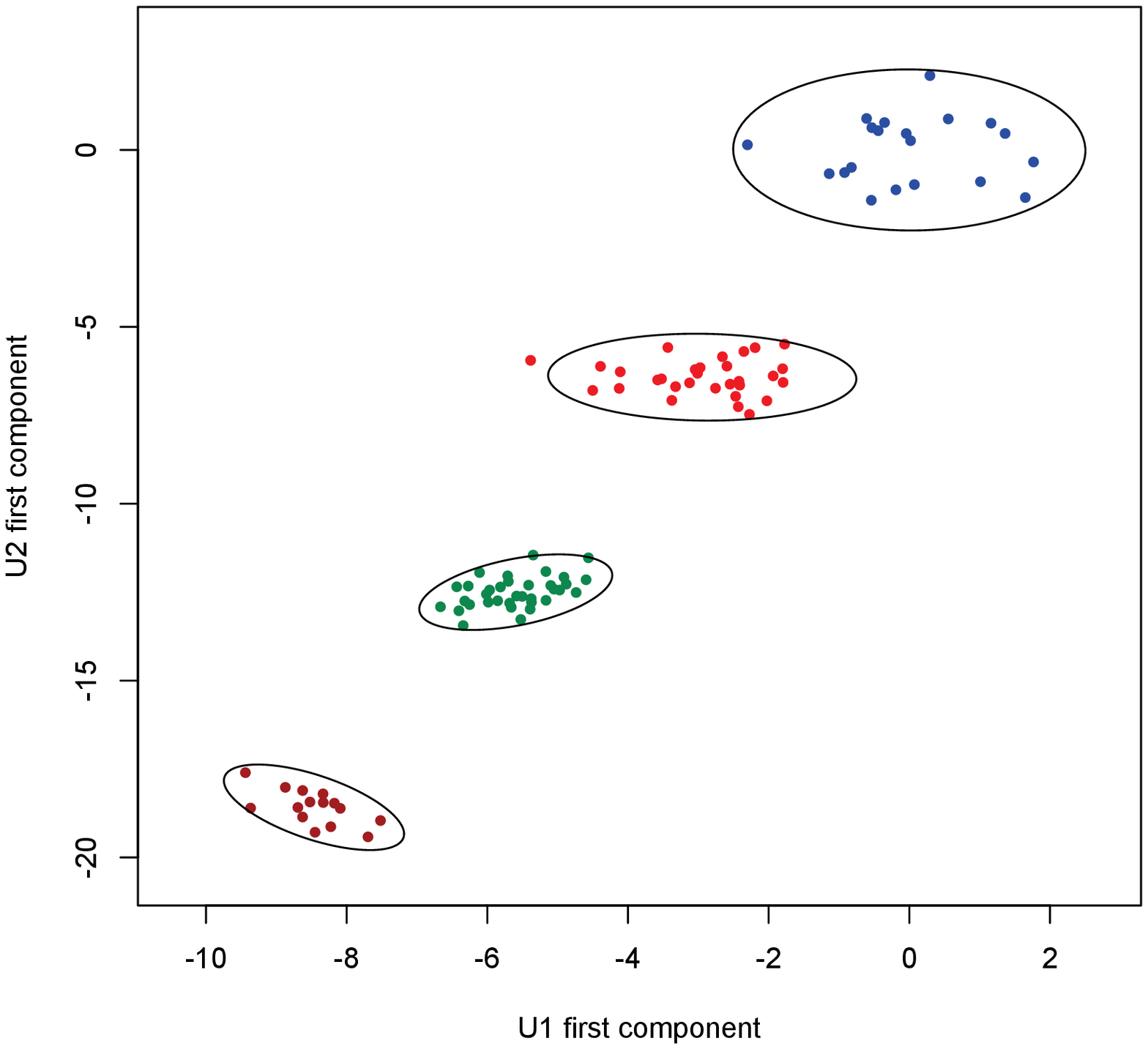} 
		\includegraphics[width=0.48\textwidth]{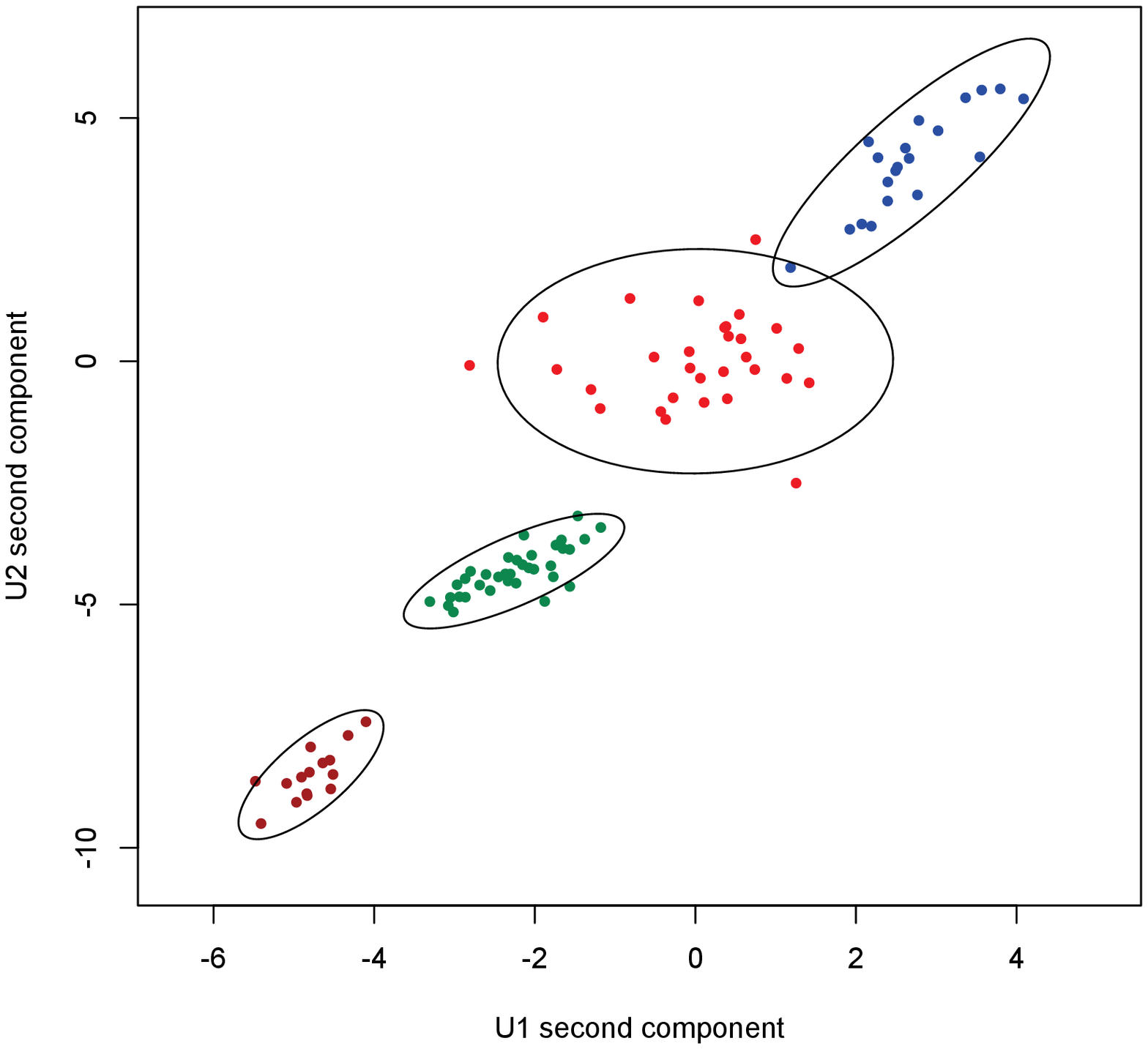} 
		\includegraphics[width=0.48\textwidth]{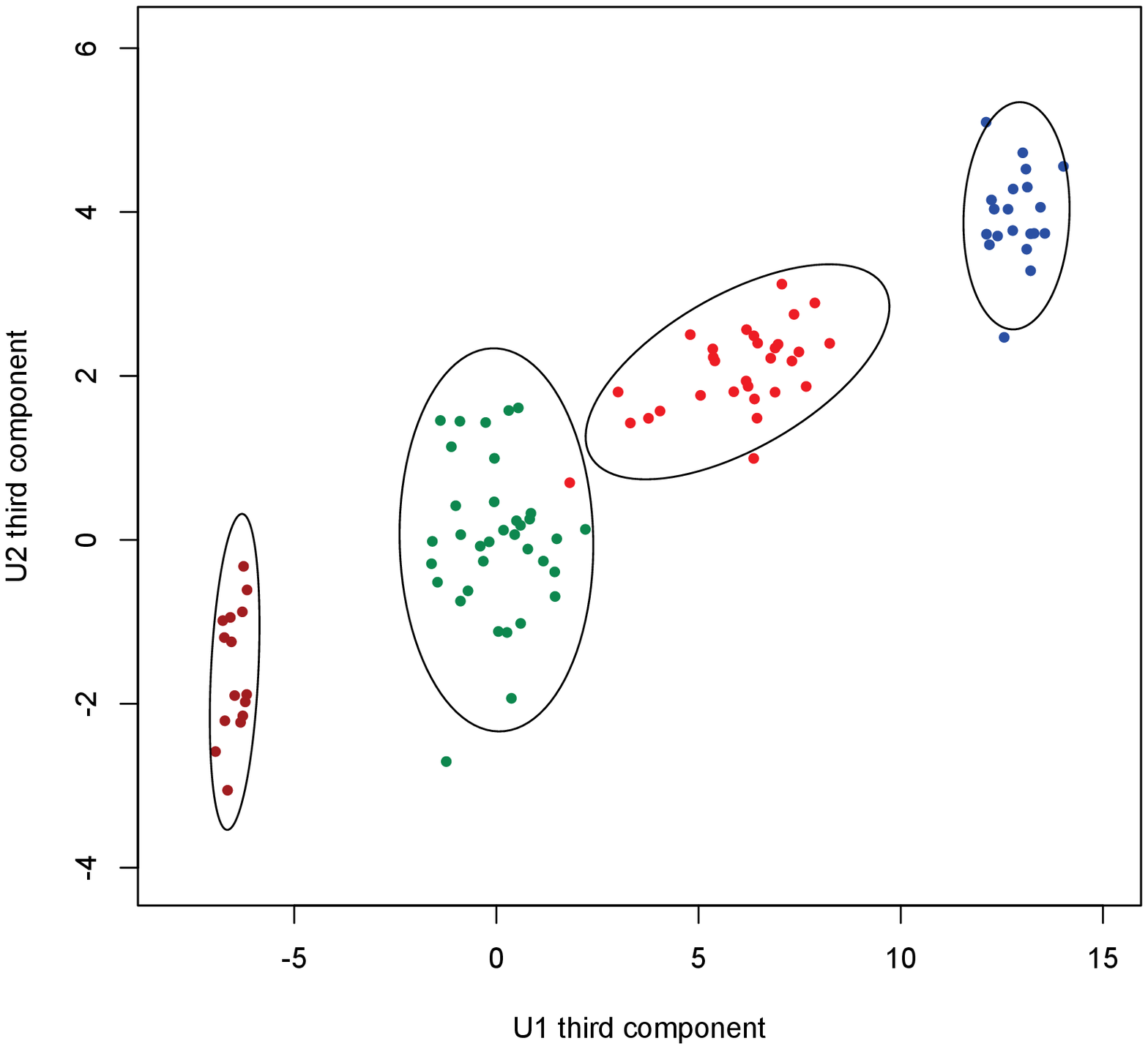} 
		\includegraphics[width=0.48\textwidth]{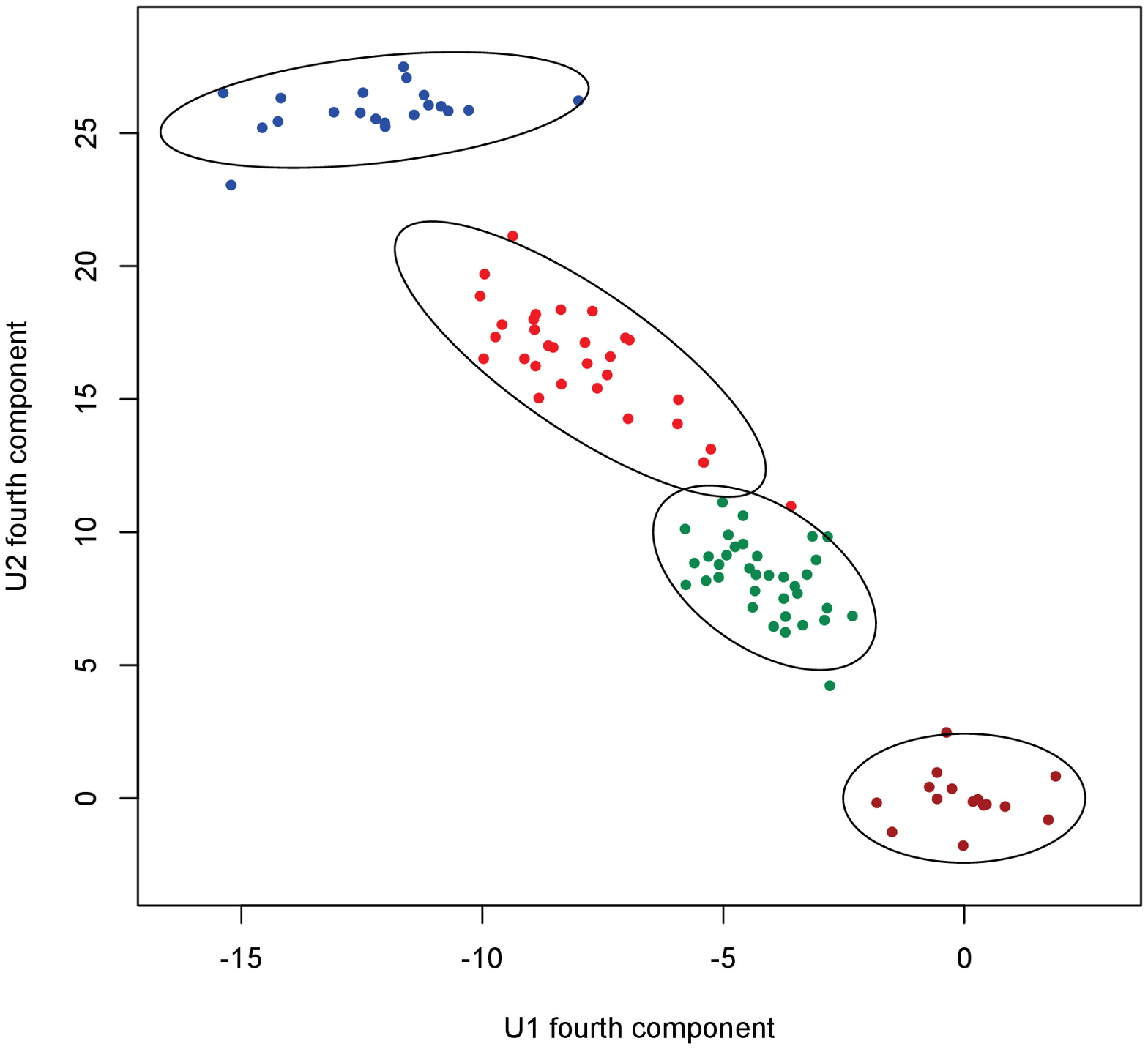} 
	\end{center}
	\caption{\rm Mixture 2: plot of the classified data on the factor spaces,  under the "right" solution given by the algorithm}
	\label{fig:Mixture_2A}
\end{figure}

\begin{figure} 
	\begin{center}
		\includegraphics[width=0.48\textwidth]{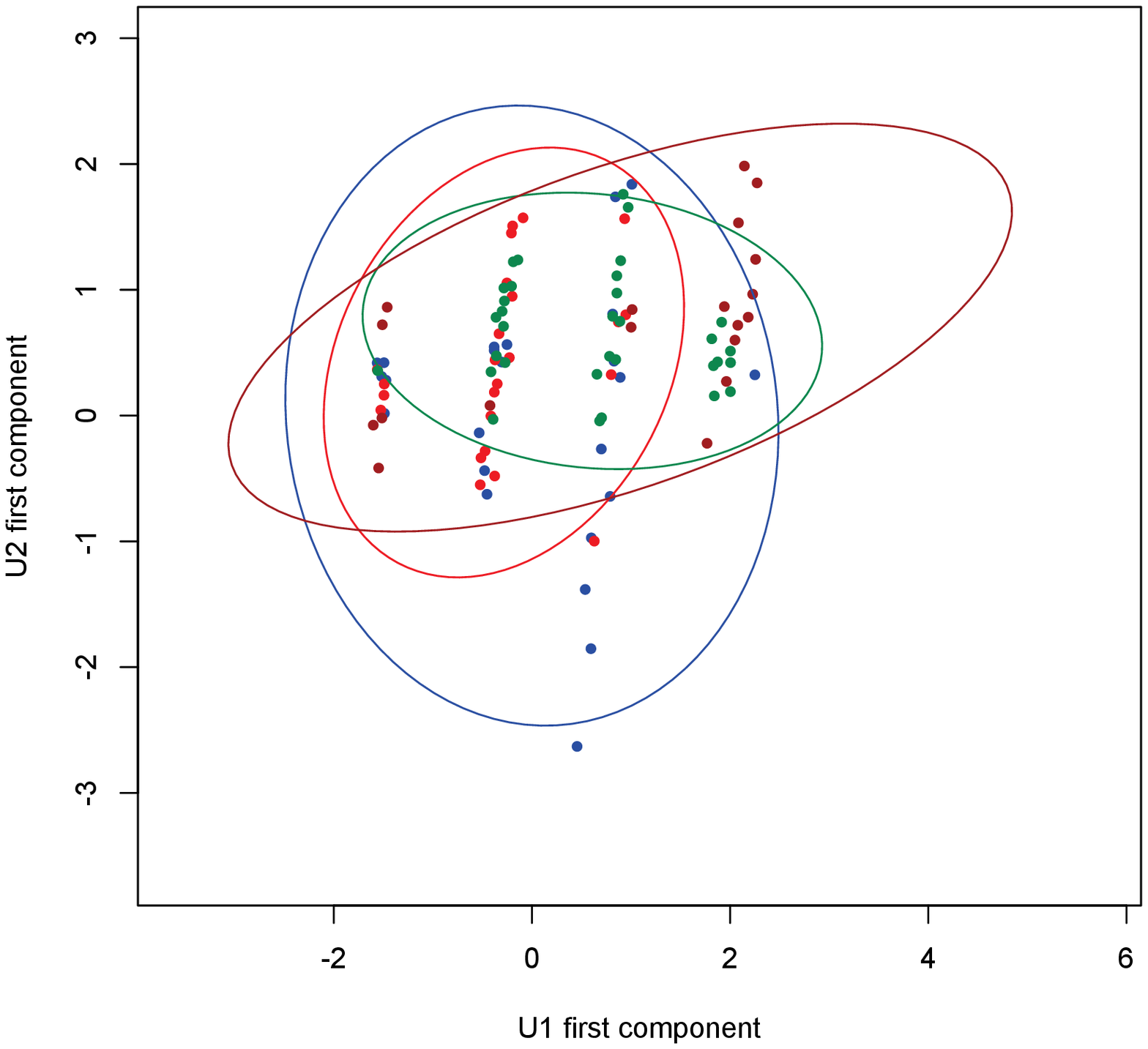} 
		\includegraphics[width=0.48\textwidth]{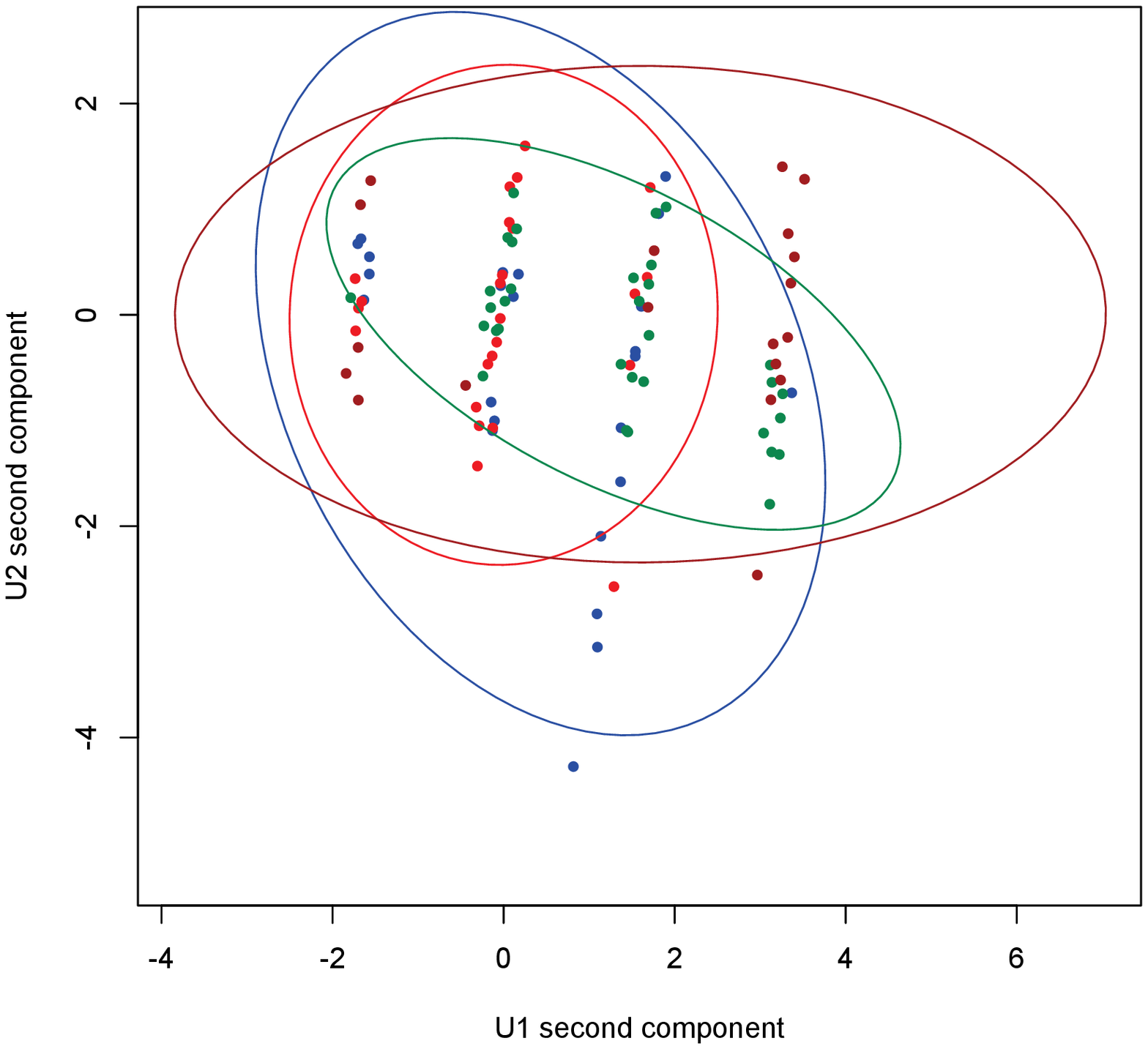} 
		\includegraphics[width=0.48\textwidth]{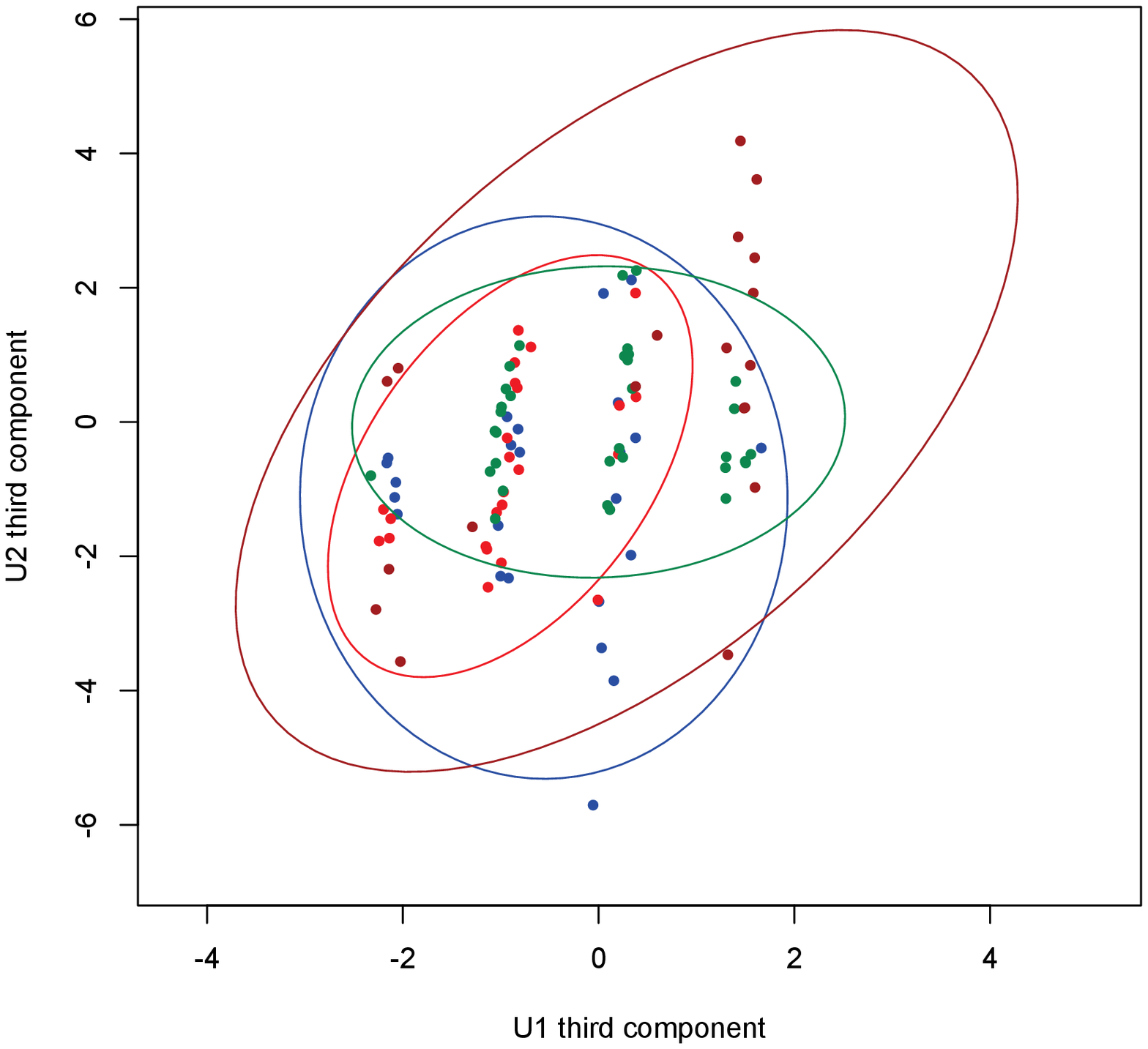} 
		\includegraphics[width=0.48\textwidth]{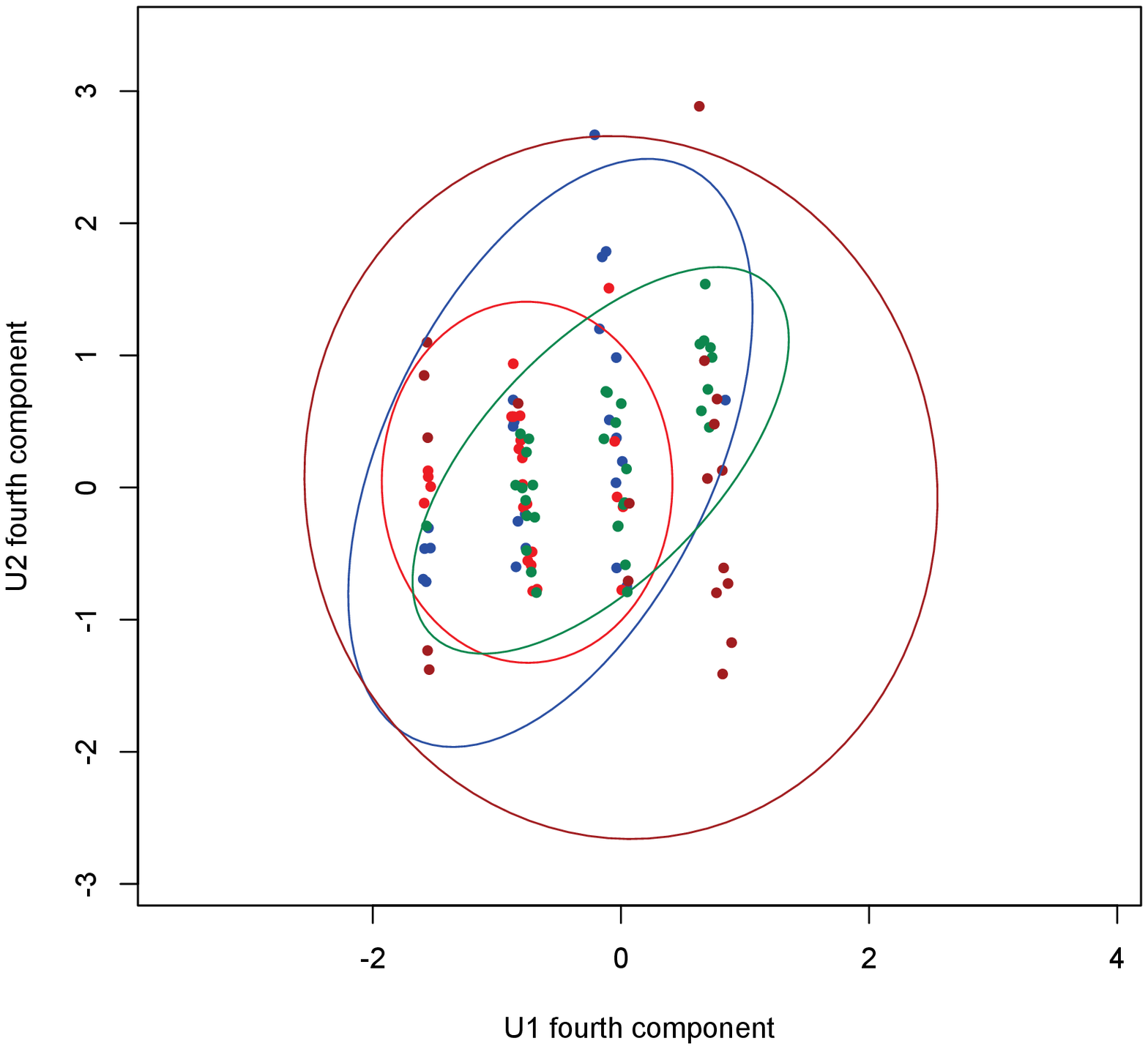} 
	\end{center}
	\caption{\rm Mixture 2: plot of the classified data on the factor spaces, giving an example of the wrong classification, which is obtained when the algorithm converges to a spurious maximum of the loglikelihood}
	\label{fig:Mixture_2B}
\end{figure}

Finally, in Figure \ref{fig:Mixture_2A} we plot the classified data on the factor spaces, under the true maximum of the likelihood function, while in Figure \ref{fig:Mixture_2B} we give the classification in some wrong factor spaces, obtained according to a spurious maximum of the likelihood function.

 \paragraph{\textsc{Mixture 3: $G=4$, $d=7$, $q=2$, $N=100$.}} \ \\
 
\begin{figure} 
	\begin{center}
		\includegraphics[width=9 cm, height=8 cm]{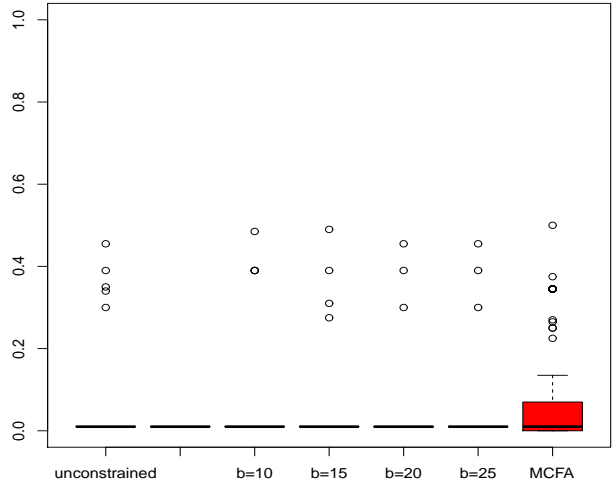} 
			\end{center}
	\caption{\rm Mixture 3: Boxplots of the misclassification error. From left to right, the first boxplot refers to the unconstrained algorithm, then the following boxplots correspond to the constrained algorithm, for $a=0.01$ and $b$ respectively set to the values  $b=6,10,15,20,25$, and finally to the MCFA algorithm.}\label{fig:BoxPlotMcLch}
\end{figure}

 The third study concerns an artificial dataset analysed in  \citet{Baek:McLa:MixtFA:2010}. It has been generated with weights $\balpha = (0.5,0.5)'$ according to the following parameters:
 \begin{align*}
\bmu_1 &= (0,0,0)'  & \bmu_2 &= (2,2,6)' 
 \end{align*}
  \vspace{-8mm}
 \begin{gather*}
\bSigma_1 = \begin{pmatrix} 4 & -1.8 &-1 \\ -1.8 & 2 & 0.9 \\  -1 & 0.9 & 2 \end{pmatrix}  \quad  \quad
\bSigma_2 = \begin{pmatrix} 4  & 1.8 & .8 \\ 1.8 & 2 & 0.5 \\   0.80 & 0.5 &2\end{pmatrix} \quad  
\end{gather*}
The covariance matrices  $\bSigma_g $ ($g=1, 2$)  have respectively the following eigenvalues:
\begin{align*}
 \lambda(\bSigma_1) &=(5.55, 1.61, 0.84)'\\
\lambda(\bSigma_2) &=(5.33, 1.73, 0.94)' 
\end{align*}
We run the unconstrained algorithm and its constrained version with the choices of $a=0.01$ and $b=6,10,15,20,25$ as before, and also we compare our proposal to the Mixture of Common Factor Analyzers (MCFA) approach of \citet{Baek:McLa:MixtCT:2011}.  The percentages of convergence to the right maximum  for the seven different cases are reported in Table \ref{tab:case3}.  
We recall that MCFA requires a common pattern between covariance matrices. This model  is greatly employed in the literature, for parsimony and to avoid potential singularities with small clusters. 
\begin{table}[h]
\begin{center}
\caption{Mixture 3: Percentage of convergence to the right maximum of the unconstrained EM, the constrained EM algorithm and the MCFA EM algorithm}\label{tab:case3}
\begin{tabular}{c ccccc c} 
\hline
unconstrained  &\multicolumn{5}{c}{constrained} & MCFA  \\
\cline{2-6}
 &$b=6$ & $b= 10$ & $b=15$ & $b=20$ & $b=25$  \\

 \hline
  95\% & 100\% & 96\% & 96\%&97\% & 97\%& 36\%\\
  \hline\hline
\end{tabular}
\end{center}
\end{table}
Over the 100 runs, the MCFA EM algorithm did not converge in 36 cases, while it always reached convergence in the other cases.  With respect to the performance of the different algorithms in terms of misclassification error, the corresponding boxplots are shown in Figure \ref{fig:BoxPlotMcLch}. We also note that the misclassification error was steadily equal to 1\% over the 100 runs for the constrained algorithm with $b=6$, it was always equal to 1\% except 5 runs for the unconstrained algorithm, while in the case of MCFA we have $Q_1=Me=1\%$, but $Q_3= 34.5\%$ and $Max=50\%$. All these results show that, to attain good performance and robustness in estimation, our proposal  works quite better. Furthermore, it allows for a  more general solution in comparison to the rigid requirement of a common pattern between covariance matrices. As a consequence, also the log-likelihood of the model obtained by our constrained algorithm ($\cL=-1032.218$) is fairly greater than the log-likelihood obtained in MCFA model ($\cL=-1147.396$).

\subsection{Real data}\label{sec:realdata}

\paragraph{\textsc{The Wine data set }}\ \\

Now we consider the wine data, proposed in \citet{Forina:etal:1986}, consisting of 
 $d=27$ chemical and physical properties of three different cultivars  of Italian wine: Barolo, Grignolino
and Barbera. This dataset is often used to test and compare the performance of various classification algorithms: among them, in \citet{McNi:Murp:Pars:2008} using parsimonious Gaussian mixture models and
in \cite{Andr:McNi:Exte:2011} using parsimonious mixtures of multivariate $t$-factor analyzers.


Consider first the complete dataset, with $d=27$. We run the EM algorithm starting from  the true classification, and using the maximum likelihood estimate $\hat{\btheta}$ we get 3 misclassified units
(i.e. Misclassification Error $1.69\%$). Based on estimates of $\bLambda_g$ and $\bPsi_g$, we get
\begin{alignat*}{3}
\lambda_{\rm max}(\hat{\bLambda}_1) & = 28513 \qquad \lambda_{\rm max}(\hat{\bLambda}_2) & = 6345 \qquad  \lambda_{\rm max}(\hat{\bLambda}_3) & = 9045 \\
\lambda_{\rm max}(\hat{\bPsi}_1) & = 27830 \qquad \lambda_{\rm max}(\hat{\bPsi}_2) & = 22532  \qquad  \lambda_{\rm max}(\hat{\bPsi}_3) & = 21573 .
\end{alignat*}
With the aim at comparing our results with the above findings in the literature, we first scaled the original data, and applied the \textit{Pgmm} package \citep{McNi:Pgmm:2011}. Using a set of three random starts, the best model (BIC) for the given range of factors and components (from 1 up to 4) is a CUU model with $q$ = 4 and $G$ = 3. The CUU acronym stands for a MGFA with patterned covariance matrices, with a common (C) volume and unconstrained (U) shapes and orientations among the $G$=3 components in the mixture. Factors for the best model are of dimension $q$=4, with  BIC= -11427.65.
The obtained classification is given by Table \ref{tab:MisclWine}, showing only 2 misclassified units.
\begin{table}[h]
\begin{center}
\caption{Pgmm package applied on the Wine dataset }\label{tab:MisclWine}
\begin{tabular}{c|ccc} 
\hline
\multicolumn{4}{c}{Classification table }  \\
\hline
 &  1 & 2&  3\\
 \hline
  1& 59  &0  &0\\
  2 & 1 &69  &1\\
  3 & 0  &0 &48  \\
  \hline
\end{tabular}
\end{center}
\end{table}

Then we employed our  approach, after scaling the data and using hierarchical clustering for initialization (as in the previously cited work). We obtained 5  misclassified units (which means a misclassification error of $2.8\%$). If we initialize the EM algorithm with the true belonging of units  and considering still 4-dimensional factors, we obtain a perfect classification. We also obtain a better fit of the model to the data, assessed by a greater penalized likelihood value, namely BIC= -10814.68, due to the lighter constraints we are imposing here.
Finally, we employed a mixture of $t$-factor analyzers, applying the teigen R-package \citep{Andr:McNi:2011}, on the scaled data. We considered patterned models, whose label is a sequence of four letters: each letter can be "C" or "U" or "I" denoting "Constrained to be equal", "Unconstrained" and "Isotropic" patterns on group covariances, and the four letters in the model label are respectively referred to  volume, shape, orientation, and the degrees of freedom of the $t$-distribution.
We got that the best fit (BIC =-11939.94) is given by CICC model with $G$=5, and this is somehow surprising as we always obtained 3 groups, by all the  methods seen so far, in particular also in the proposed constrained EM approach for gaussian factors.

\paragraph{\textsc{The Flea Beetles data set }}\ \\

The flea beetles data were introduced by \citet{Lubi:DiTa:1962} and are available within the GGobi software, see \citet{Sway:Cook:GGob:2006}. 
Data were collected on $74$ specimens  of flea beetle of the genus \textit{Chaetocnema}, which contains three species:  {\em concinna}, {\em heptapotamica}, or {\em heikertingeri}. Measurements were collected on the width (in the fore-part and from the side) and angle  of the aedeagus, on the width of the first and second joint of the tarsus, and on the width of the head between the external edges of the eyes of each beetle.

The goal of the original study was to form a classification rule to distinguish the three species.
To this aim, we considered $q=2$ factors, according to the results of \citet{Andr:McNi:Exte:2011}, and
we run firstly the unconstrained algorithms. Over the 100 runs, the unconstrained EM algorithm never reached the true solution, and summary statistics (minimum, first quartile $Q_1$, median $Q_2$, third quartile
$Q_3$ and maximum) about  the distribution of the misclassification error over the 100 runs are reported in Table \ref{tab:MisclassFlea}.
\begin{table}[h]
\begin{center}
\caption{Flea Beetles data: Summary  statistics of the distribution of the Misclassification Error over 100 runs of the unconstrained EM algorithm}\label{tab:MisclassFlea}
\begin{tabular}{ccccc} \hline
  \multicolumn{5}{c}{Misclassification Error}  \\
  min &  $Q_1$& $Q_2$ & $Q_3$ & max \\ \hline
  4.1\% & 28.0\% & 36.5\% & 41.9\% & 51.4\% \\
  \hline\hline
\end{tabular}
\end{center}
\end{table}
The first results motivated us to run also the constrained EM algorithm, to see if it improves convergence to the right maximum and consequent classification. Tacking into account that \[ \min_{i,g} \lambda_i(\bSigma_g)=0.64 \qquad \max_{i,g} \lambda_i(\bSigma_g)=191.55,\] we considered constrained estimation
with 
\begin{center} lower bound $a$  either 0.1 or 0.5,  $\quad \quad$ and $\quad \quad$
upper bound $b$ either 200 or 300.\\
\end{center}

Over the 100 runs, the constrained algorithm steadily improves all results, as it can be seen in Table \ref{tab:Flea}, which shows also that the best results can be obtained with the tightest constraints, i.e. $a=0.05, b= 200$.
\begin{table}[h]
\begin{center}
\caption{Flea Beetles data: Percentage of convergence to the right maximum of the unconstrained EM and  the constrained EM algorithm}\label{tab:Flea}
\begin{tabular}{c cccc} 
\hline
unconstrained  &\multicolumn{4}{c}{constrained}  \\
\cline{2-5}
 &$a=0.1, b=200$ & $a=0.05, b= 200$ & $a=0.1, b=300$ & $a=0.5, b=300$  \\

 \hline
  0\% & 31\% & 34\% & 21\%&17\% \\
  \hline\hline
\end{tabular}
\end{center}
\end{table}

\begin{figure} 
	\begin{center}
		\includegraphics[width=9 cm, height=8 cm]{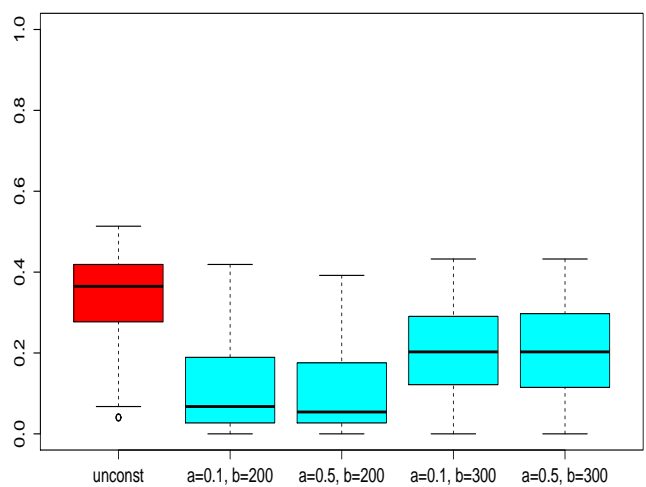} 
			\end{center}
	\caption{\rm Flea Beetles data: Boxplots of the misclassification error. From left to right, the first boxplot refers to the unconstrained algorithm, then the following boxplots correspond to the constrained algorithm, for each pair of  bounds $(a,b)$ }\label{fig:BoxPlotFlea}
\end{figure}

Figure \ref{fig:BoxPlotFlea}  provides the boxplots of the distribution of the of the $100$ misclassification errors in the  sequences of $100$ runs for both unconstrained and constrained algorithms. The impact of the lower bound $a$ on the estimation is  critical, but it seems not to depend too much on its value (remember that its purpose is to protect against divergence of the algorithm) while the upper bound $b$ crucially drives the classification results, showing the best performance when it mimics the value of the largest eigenvalue of the $\bSigma_g$'s. 
As a final comment, it is worth mentioning that, when dealing with  EM estimation based on random starts, authors in the literature usually give results in terms of "best outcome over a small number of runs", say 10 runs for instance. Therefore, we can conclude that the constrained algorithm (having a performance of $31\%$)  provides the true solution and the perfect classification for the Flea Bleetles dataset.

\section{Concluding remarks}\label{sec:concluding}
Mixtures of factor analyzers are  commonly used to explain data, in particular, correlation between variables in multivariate observations, allowing also for dimensionality reduction. For these models, as well as for gaussian mixtures, however, the loglikelihood function may present spurious  maxima and  singularities and this is due to specific patterns of the estimated covariance structure.  It is known, from the literature, that a constrained formulation of the EM algorithm considerably reduces such drawbacks for gaussian mixtures. Motivated by these considerations,  in this paper we introduced a constrained approach for  gaussian mixtures of factor analyzers. In particular we implemented a methodology to  maximize the likelihood function in a constrained parameter space, having no singularities and a reduced number of spurious local maxima. The performance of the newly introduced estimation approach has been shown  and compared to the usual non-constrained one, as well as to the approach based on common factors. To this purpose  we present numerical simulations on synthetic samples and  applications to real data sets widely employed in the literature. The results shows that the problematic convergence of the EM, even more critical when dealing with factor analyzers, can be greatly improved.

\end{document}